\documentclass[12pt, draftclsnofoot, onecolumn]{IEEEtran}

\usepackage{epsfig,makeidx,color,mathtools,bigints,graphicx,amsbsy,amsmath,amssymb,euscript,verbatim}
\usepackage{url}

\DeclareMathOperator{\argmax}{argmax}



\newtheorem{lemma}{Lemma}
\newtheorem{prop}{Proposition}
\newtheorem{definition}{Definition}

\newtheorem{corollary}{Corollary}

\newtheorem{theorem}{Theorem}

\newcommand{\Z}{\mathbb{Z}} 
\newcommand{\R}{\mathbb{R}} 
\newcommand{\C}{\mathbb{C}} 

\newcommand{\diag}{{\hbox{diag}}}
\renewcommand{\det}{{\hbox{\rm det}}}

\begin{document}

\title{Locally Diverse Constellations from the Special Orthogonal Group}

\author{David A. Karpuk, \IEEEmembership{Member, IEEE,}
        Camilla Hollanti, \IEEEmembership{Member, IEEE}
\thanks{D. Karpuk and C. Hollanti are with the Department of Mathematics and Systems Analysis, Aalto University, Espoo, Finland.  emails: \{david.karpuk, camilla.hollanti\}@aalto.fi}
\thanks{D. Karpuk is supported by Academy of Finland Postdoctoral grant \#268364.}
\thanks{C. Hollanti is supported by Academy of Finland grants \#276031, \#282938, and \#283262, and by a grant from Magnus Ehrnrooth Foundation, Finland. The support from the European Science Foundation under the COST Action IC1104 is also gratefully acknowledged.}
}
\maketitle

\IEEEpeerreviewmaketitle

\begin{abstract}

To optimize rotated, multidimensional constellations over a single-input, single-output Rayleigh fading channel, a family of rotation matrices is constructed for all dimensions which are a power of 2. This family is a one-parameter subgroup of the group of rotation matrices, and is located using a gradient descent scheme on this Lie group. The parameter defining the family is chosen to optimize the cutoff rate of the constellation. The optimal rotation parameter is computed explicitly for low signal-to-noise ratios.

These rotations outperform full-diversity algebraic rotations in terms of cutoff rate \textcolor{black}{at low SNR (signal-to-noise ratio)} and \textcolor{black}{bit error rate at high SNR} \textcolor{black}{in dimension $n = 4$}. However, a QAM \textcolor{black}{(quadrature amplitude modulation)} constellation rotated by such a matrix lacks full diversity,  in contrast with the conventional wisdom that good signal sets exhibit full diversity.  \textcolor{black}{A new notion of diversity, referred to as local diversity, is introduced to attempt to account for this behavior.} Roughly, a locally fully diverse constellation is fully diverse only in small neighborhoods. A local variant of the minimum product distance is also introduced and is shown experimentally to be a superior predictor of constellation performance than the minimum product distance \textcolor{black}{in dimension $n = 4$}.


\end{abstract}

\begin{IEEEkeywords}
Rayleigh fading channel, rotated constellations, non-uniform constellations, capacity, cutoff rate, Lie groups, full diversity
\end{IEEEkeywords}

\IEEEpeerreviewmaketitle

\section{Introduction}

\subsection{Rotated Constellations and the Cutoff Rate}

Communication over a wireless channel typically involves transmitting a signal which is subsequently affected by fading and noise. Ensuring that such transmission is done reliably requires an energy-efficient constellation resistant to these effects. It is known that rotating constellations is an effective way to combat the effect of fading, while maintaining the energy of each codeword.  Rotations effectively combat fading by increasing the diversity of the constellation.

Numerous algebraic and number-theoretic techniques exist to construct fully-diverse \textcolor{black}{rotated cubic} lattices with good minimum product distance \cite{boutros,boutrousgood,OV}, of which a finite subset is selected to serve as the constellation.  These algebraic methods are effective at producing fully-diverse constellations which have good performance in terms of error rate.  However, the rigidity of the algebraic constructions imposes a lack of flexibility in adjusting the particular rotation matrix in the presence of, for example, an outer error-correcting code or varying SNR (signal-to-noise ratio).  Furthermore, the number-theoretic constructions of \cite{OV} assume high SNR and that the signal set is essentially infinite, that is, consists of an entire lattice in $\R^n$.  

We investigate potential improvements in the algebraic method of \cite{OV} by dropping these assumptions and investigating rotations via a finer and SNR-dependent objective function controlling constellation performance.  This objective function is an approximation of the capacity of the constellation known as the \emph{cutoff rate}, \textcolor{black}{which has a long history of being used as design criterion, but almost exclusively for erasure and AWGN channels; see the classical references \cite{massey, forney_ung} and the references therein.  It has been utilized much less for the fading channel model we consider, in favor of coarser, more algebraic measures of constellation performance.}

Rotating constellations to improve CM (coded modulation) capacity has been investigated previously in \cite{fabregas,herath,hero}.  Our work is partially inspired by \cite{herath}, in which the authors approximate the CM capacity by the cutoff rate, and use the latter to construct good rotation matrices for $4$-QAM and $16$-QAM (quadrature amplitude modulation) constellations in $\C^4$ and $\C^6$.  The cutoff rate was also considered in conjunction with constellation design in \cite{hero,boulle2}.  \textcolor{black}{Our goal is to apply the general method of \cite{herath} of considering rotations which attempt to maximize the cutoff rate to the channel model of \cite{OV}, for which one often constructs algebraic rotations.}  The cutoff rate for the channel we consider is a simple rational function of the coordinates of the constellation points and therefore, unlike working with the CM capacity directly, does not require one to numerically compute \textcolor{black}{numerically unstable} unbounded integrals during optimization.  In particular, the cutoff rate is simple enough to be susceptible to the methods of \cite{edelman} for numerical optimization over rotation matrices, \textcolor{black}{which we employ to  investigate optimal rotation matrices}.

\subsection{Non-Uniform Constellations}

\textcolor{black}{By a \emph{non-uniform} constellation, one typically means a ``perturbation'' of a standard QAM constellation, which heuristically behaves like a closer approximation of a Gaussian distribution than the standard ``square'' QAM constellations.  Non-uniform constellations for the AWGN channel have been studied in \cite{zoellner}, wherein the authors perform SNR-dependent optimization of such constellations, and in \cite{hossain} with the goal of optimizing the BICM (bit interleaved coded modulation) capacity.  Non-uniform constellations have been shown to have good error rate performance for the Gaussian \cite{betts} and Rayleigh \cite{zhongtakahara} channels.  Furthermore, non-uniform $64$-QAM and $256$-QAM constellations have been considered in DVB-NGH (Digital Video Broadcasting - Next Generation Handheld) implementation \cite{DVBNGH} due to the improvement in capacity, and are under consideration for inclusion in the ATSC (Advanced Television Systems Committee) 3.0 industry standard \cite{lachlan}.}

On the other hand, the DVB consortium has established rotated QAM constellations as a part of the DVB-T2 industry standard \cite{DVBT2}, and rotated QAM constellations in $\R^2$ and $\R^4$ as part of the DVB-NGH  (next-generation handheld) standard \cite{DVBNGH}.  \textcolor{black}{A concise summary of such modulation techniques can be found in \cite[\S 11]{big_broadcasting_book}.}  A natural question is whether combining non-uniformity and rotations can offer an additional increase in capacity, and thus one would like to know how to find good rotations for arbitrary constellations, not just traditional $M$-QAM.  \textcolor{black}{As optimal non-uniform constellations can change with the SNR, and our goal is partly to rotate such constellations, we seek a rotation method which is SNR-dependent as well.  Working with the CM capacity and cutoff rate also demands that one considers SNR-dependent rotations, since both of these objective functions depend on the SNR.  As we shall see in Section \ref{nuqam2d}, optimal rotations can be highly SNR-dependent and optimizing rotations with respect to SNR can improve error rate performance}.  


\subsection{Summary of Main Contributions}

\textcolor{black}{We present a general method for constructing good rotations of arbitrary constellations for single-input, single-output Rayleigh fast fading channels in dimensions which are a power of 2.}  Constructing rotations in arbitrarily high dimensions means abandoning explicit parameterizations of such matrices, since such parameterizations become non-canonical and unwieldy as the dimension of the ambient space increases.  The collection of all rotations of $\R^n$ is the \emph{special orthogonal group} $SO(n)$ which has dimension $(n^2-n)/2$ as a real manifold, meaning any parameterization of $n\times n$ rotation matrices requires at least $(n^2-n)/2$ variables.  We will present the necessary mathematical framework in Section \ref{graddescent}. 

Our approach to constructing optimal rotation matrices differs largely from previous work on the subject, in that we effectively replace explicit parameterizations of rotation matrices with the matrix exponential map
\begin{equation}
\exp: \frak{so}(n) \rightarrow SO(n)
\end{equation}
where $\frak{so}(n)$ is the Lie algebra of all skew-symmetric matrices.  The general mathematical framework of Lie groups and Lie algebras allows us to construct well-performing families ${\bf{Q}}_{2^k}(t)$ of $2^k$-dimensional rotation matrices for all $k$, which we view as one-parameter subgroups of $SO(2^k)$.  The problem of optimizing over all $(n^2-n)/2$ parameters defining a rotation matrix is thus reduced to optimizing over just a single parameter $t\in[0,\pi/2]$, which is easily done by exhaustive search and can be catered to the given SNR and constellation.  

The flexibility of the parameter $t$ is also beneficial if one wishes to further optimize due to the presence of an outer error-correcting code, a particular decoding algorithm, a particular bit interleaver, etc.  In fact this is essentially what is done in the DVB-NGH standard \cite[\S6]{DVBNGH}, which considers the (transpose of the) same family of matrices ${\bf{Q}}_4(t)$ which we construct explicitly, and optimizes the parameter $t$ to suit the outer LDPC code rate and modulation order.  By generalizing this construction to any dimension which is a power of $2$, we hope to provide robust and adaptive families of rotation matrices to suit a number of applications.  \textcolor{black}{In general, we view our method as an alternative to the algebraic and number-theoretic constructions which accounts for the finiteness of the SNR, the boundedness of the constellation, and is adaptable to a number of potential objective functions which control system performance.  We hope to generalize our method to the MIMO setting, and analogously improve upon the performance of division algebra codes for MIMO Rayleigh fading channels \cite{OBV}, lattice codes for the Rician fading channel \cite{amin_rician}, and full-diversity unitary precoding techniques for MIMO channels \cite{amin_full_diversity}}.  Our adaptive approach may also be useful in newer protocols such as SCMA (sparse code multiple access) \cite{scam}, in which algebraic rotations are proposed, but the complexity of SCMA codebook design may demand more intricate constructions in the future.

This article is intended to summarize, expand upon, and provide a more satisfying theoretical framework for the results of the conference papers \cite{karpukrot,karpukrot2} by the current authors.  In particular, the family of rotation matrices ${\bf Q}_{2^k}(t)$ we consider was discussed in \cite{karpukrot2} but we present here a simple, explicit formula in Section \ref{explicit_construction} for their construction.  We find in Section \ref{lowsnr} the optimal rotation within the family ${\bf Q}_{2^k}(t)$ in the low SNR regime and confirm our results via experiment.  We investigate in Section \ref{diversity} the diversity order of our rotated constellations and show in Section \ref{n4results} that despite the lack of full diversity, they outperform the fully-diverse algebraically-rotated constellations of \cite{OV} in terms of both cutoff rate \textcolor{black}{(at low SNR)} and error performance \textcolor{black}{(at high SNR)} \textcolor{black}{in dimension $2^k = 4$}.  We attribute this to a property we call \emph{local full diversity}, which describes constellations which may not be fully diverse, but are so if one restricts to small neighborhoods.  Our experiments indicate that constellations which enjoy this less stringent property can offer modest gains over fully diverse signal sets.  This is in contrast with the conventional wisdom presented in \cite{OV} that optimal signal sets have full diversity.




\section{Coded Modulation Capacity and Cutoff Rate}

We consider the general problem of constructing good constellations for Rayleigh fast fading single-input single-output (SISO) channels for transmission over $n$ time instances.  We assume perfect channel state information at the receiver, and as in \cite[\S2]{OV}, \textcolor{black}{by performing phase cancelation at the receiver}, separating real and imaginary parts, and employing a bit interleaver, we can model the channel as
\begin{equation}\label{model}
\bf{y} = \bf{h}\bf{x} + \bf{z}
\end{equation}
where 
\begin{itemize}
\item[$\bullet$] ${\bf x} = (x_1,\ldots,x_n)^t \in \R^n$ is a column vector of transmitted data, selected uniformly at random from a finite constellation $\mathcal{X}\subset \R^n$ \textcolor{black}{of size $|\mathcal{X}| = 2^q$,}
\item[$\bullet$] ${\bf h}=\diag(h_1,\ldots,h_n)$ is a real diagonal $n\times n$ matrix with $h_i$ a Rayleigh random variable \textcolor{black}{(i.e.\ the norm of a circularly symmetric complex Gaussian random variable)} with $\mathbf{E}(h_i^2) = 1$,
\item[$\bullet$] ${\bf z} =(z_1,\ldots,z_n)^t \in\R^n$ is a noise vector with $z_i$ a real zero-mean Gaussian random variable with variance $N_0$, and
\item[$\bullet$] ${\bf y} \in \R^n$ is the received vector.
\end{itemize}
We assume that the constellation is subject to an energy constraint of the form
\begin{equation}
\mathbf{E}||{\bf x}||^2 = \frac{1}{|\mathcal{X}|}\sum_{{\bf x}\in\mathcal{X}}||{\bf x}||^2 = P
\end{equation}
for some fixed $P$, where $|\mathcal{X}|$ denotes the cardinality of $\mathcal{X}$.  Let us define the average energy per bit $E_b$ by $P/\log_2(|\mathcal{X}|)$.  We will present simulation results in terms of $E_b/N_0$ in decibels, but we will slightly abuse terminology and occasionally refer to this quantity as the SNR (signal-to-noise ratio) as well.

If ${\bf T}$ is any $n\times n$ \textcolor{black}{real} matrix (in all cases we consider ${\bf T}$ will either be a channel matrix or a rotation matrix), then by the notation ${\bf T}\mathcal{X}$ we mean the constellation
\begin{equation}
{\bf T}\mathcal{X} = \{{\bf Tx}\ |\ {\bf x}\in\mathcal{X}\}\subset\R^n.
\end{equation}

\subsection{Definitions and Basic Notions}

In this section we recall some familiar formulas for the CM capacity of an $n$-dimensional constellation $\mathcal{X}\subset\R^n$ \textcolor{black}{of size $|\mathcal{X}| = 2^q$}.  This discussion follows that of \cite{herath}, but for the slightly different channel model (\ref{model}).  If we set ${\bf h} = {\bf I}_n$ in (\ref{model}), the mutual information of the output $\mathcal{Y}$ and the constellation $\mathcal{X}$ is
\begin{equation}
\begin{aligned}
\textcolor{black}{I(\mathcal{Y},\mathcal{X})} = \log_2(|\mathcal{X}|) - \sum_{{\bf x} \in \mathcal{X}} \bigintssss_{\R^n}&\frac{\exp\left(-||{\bf y}-{\bf x}||^2/N_0\right)}{|\mathcal{X}|(\pi N_0)^{\frac{n}{2}}}   \\
&\times \log_2\left[\sum_{\substack{{\bf x}'\in\mathcal{X} \\ {\bf x}' \neq {\bf x}}}\exp\left(\frac{||{\bf y}-{\bf x}||^2-||{\bf y} - {\bf x}'||^2}{N_0}\right)\right]\ d{\bf y}
\end{aligned}
\end{equation}
If ${\bf h} = \text{diag}(h_1,\ldots,h_n)$ is a fixed fading matrix, then the conditional mutual information of $\mathcal{Y}$ and $\mathcal{X}$ given ${\bf h}$ is 
\begin{equation}
\textcolor{black}{I(\mathcal{Y},\mathcal{X};{\bf h}) = I(\mathcal{Y},{\bf h}\mathcal{X}).}
\end{equation}
One defines the \emph{coded modulation capacity} (CM capacity) of $\mathcal{X}$ by taking the expectation over all ${\bf h}$:
\begin{equation}
C^{\text{CM}}(\mathcal{X}) = \mathbf{E}_{\bf h} I(\mathcal{Y},\mathcal{X};{\bf h})
\end{equation}
For fixed $n$ and $E_b/N_0$, one would ultimately like to optimize $C^{\text{CM}}(\mathcal{X})$ over all constellations $\mathcal{X}$, but the large amount of numerical integration required to compute $C^{\text{CM}}(\mathcal{X})$ makes this intractable, especially with increasing dimension and constellation size.  \textcolor{black}{In fact, even evaluating $C^{\text{CM}}(\mathcal{X})$ with the numerical precision required to apply our optimization techniques seems out of reach, even for small constellations in small dimensions.}  Our first step is therefore to replace $C^{\text{CM}}(\mathcal{X})$ with a more tractable objective function.

\subsection{The Cutoff Rate}

It is known \cite{herath,hero,caire} that one can bound $I(\mathcal{Y},\mathcal{X};{\bf h})$ below by a quantity $R_0(\mathcal{X};{\bf h})$, which for the model (\ref{model}) is given by
\begin{equation}
R_0(\mathcal{X};{\bf h}) = q - \log_2\left[1+2^{-q}\sum_{{\bf x}\neq {\bf y}\in \mathcal{X}} \exp\left(\frac{-||{\bf h}({\bf x}-{\bf y})||^2}{8N_0}\right)\right]
\end{equation}
\textcolor{black}{where recall that $|\mathcal{X}| = 2^q$, so that each constellation point encodes $q$ bits.  One can now define
\begin{equation}
\textcolor{black}{R_0(\mathcal{X}) = \mathbf{E}_{\bf h} R_0(\mathcal{X};{\bf h})}
\end{equation}
and it follows trivially that $C^{\text{CM}}(\mathcal{X})\geq R_0(\mathcal{X})$.  }

To establish numerical techniques which are independent of \textcolor{black}{computing an expectation over all ${\bf h}$}, one uses Jensen's inequality to further lower bound $\textcolor{black}{R_0(\mathcal{X})}$ by the \emph{cutoff rate} $R(\mathcal{X})$, defined by
\begin{equation}\label{cutoff_rate}
R(\mathcal{X}) = q - \log_2\left[1+2^{-q}\sum_{{\bf x}\neq {\bf y}\in \mathcal{X}}\prod_{i=1}^n\frac{1}{1+\frac{(x_i-y_i)^2}{8N_0}}\right]
\end{equation}
For some fixed ${\bf x} \neq {\bf y}\in\mathcal{X}$, the corresponding summand in (\ref{cutoff_rate}) is related to the pairwise error probability of confusing ${\bf y}$ for ${\bf x}$ at the receiver \cite{boutrousgood}.  Our goal is now to optimize our constellation $\mathcal{X}$ by treating $R(\mathcal{X})$ as an objective function we wish to optimize for a given $E_b/N_0$, dimension $n$, and modulation order $|\mathcal{X}|$.  We briefly remark that most authors \cite{herath,caire} use the term ``cutoff rate'' to refer to the quantity $\textcolor{black}{R_0(\mathcal{X})}$, but we reserve the term for $R(\mathcal{X})$ to simplify terminology, since it is the main objective function we consider.  

Studying the cutoff rate eliminates the need to perform the numerical integration necessary to evaluate $C^{\text{CM}}(\mathcal{X})$, which can be computationally demanding \textcolor{black}{or even infeasible}, especially since the integral is unbounded \textcolor{black}{and the integrand is numerically unstable.  Furthermore, the use of Jensen's Inequality has eliminated computing ${\bf E_h}(\cdot)$, which would further require computing a second numerically unstable integral, or measuring $R_0(\mathcal{X})$ via Monte Carlo simulations. Even small improvements in $C^{\text{CM}}(\mathcal{X})$, $R_0(\mathcal{X})$, and $R(\mathcal{X})$ can introduce non-trivial benefits in terms of, for example, error performance, and small gains in the former two quantities are necessarily very difficult to measure, especially for larger dimensions and modulation orders.  Hence we will almost exclusively measure performance in terms of $R(\mathcal{X})$. }  Note that the dependence of $R(\mathcal{X})$ on the noise variance $N_0$ allows us to tailor our constellations to particular SNR regimes.  


\section{Non-Uniform QAM Constellations in $\R^2$}\label{nuqam2d}

Let us begin by studying the model (\ref{model}) when $n = 2$.  The purpose of this section is to introduce non-uniform constellations and demonstrate their utility.  Non-uniform QAM (NUQAM) constellations have been included in the DVB-NGH standard \cite[\S6]{DVBNGH} for modulation orders of $M = 64, 256$, wherein their defining parameters have been optimized according to the rate of an outer LDPC code.  Here we demonstrate how optimal rotations of constellations vary when one changes both the SNR and the non-uniformity parameters, which helps motivate our more intricate constructions in later sections when we increase the value of $n$.

\subsection{Optimal Non-Uniform Constellations}

Let $q\geq 4$ and consider a set of parameters $\alpha = (\alpha_1,\ldots,\alpha_{2^{q/2 - 1}})\in\R^{2^{q/2 - 1}}$ with $\alpha_i>0$ for all $i$.  By an \emph{$M$-NUQAM constellation} $\mathcal{X}(\alpha)\subset\R^2$ we mean the direct product of the set 
\begin{equation}
\{-\alpha_{2^{q/2 - 1}},\ldots,-\alpha_1,\alpha_1,\ldots,\alpha_{2^{q/2 - 1}}\}
\end{equation} 
with itself, such that $|\mathcal{X}(\alpha)| = M$.  For example, when $q=4$ and $\alpha = (1,4)$, we have the $16$-NUQAM constellation
\begin{equation}
\mathcal{X}(1,4) = \{(\pm1,\pm1),(\pm1,\pm4),(\pm4,\pm1),(\pm4,\pm4)\}.
\end{equation}
The $\alpha_i$ should be interpreted as perturbed versions of the normal parameters defining the standard $M$-QAM constellations.  
\begin{table}[h!] 
\begin{center}
\caption{Some Example $M$-NUQAM Constellations}\label{nuqam_table}
\begin{tabular}{|c|c|c|}
\hline
$E_b/N_0$ (dB) & $M$ & optimal $\alpha = (\alpha_1,\ldots,a_{2^{q/2-1}})$ for $M$-NUQAM\\ \hline
8 & 16 & (0.9732, 3.0088) \\ \hline
12 & 64 & (0.9179, 2.7927, 4.8112, 7.2257) \\ \hline
15 & 256 & (0.8912, 2.6844, 4.5119, 6.4022, 8.3956, 10.5573, 13.0147, 16.1037) \\ \hline
\end{tabular}
\end{center}
\end{table}
We will refer to the $\alpha_i$ as the \emph{non-uniformity parameters}.  Such constellations are known to increase capacity at lower SNR ranges \cite{DVBNGH}.  

\begin{figure}[h!]
\centering
\hspace{-2em}\includegraphics[width = .4\textwidth]{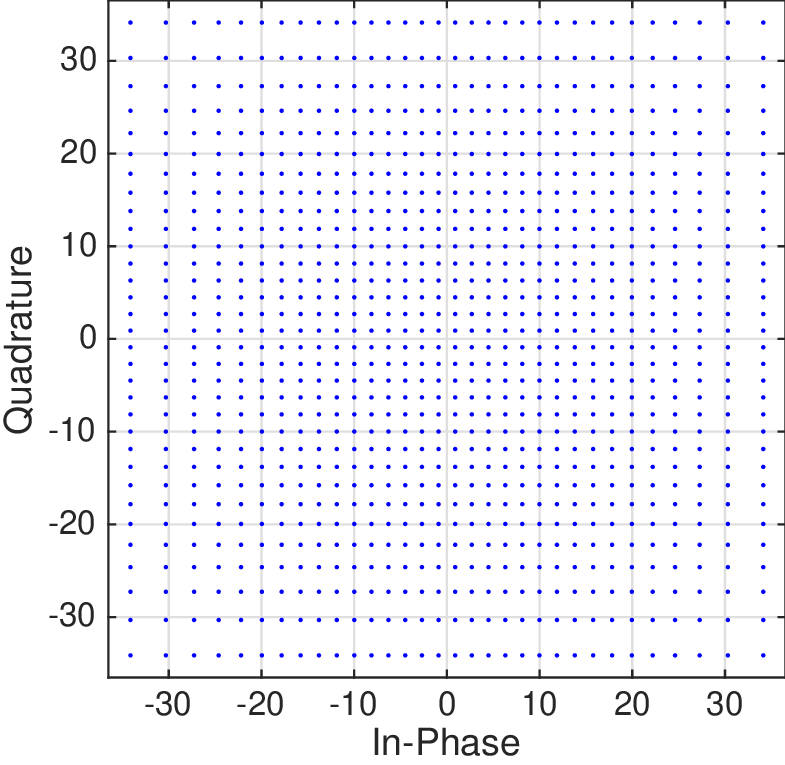}
\caption{Scatter plot of a 1024-NUQAM constellation in $\R^2$, optimized for $E_b/N_0 = 20$ dB.}\label{nuqam_scatter}
\end{figure}

For a fixed constellation size and $E_b/N_0$, we optimize the non-uniformity parameters $\alpha$ by performing steepest descent on the objective function $R(\mathcal{X}(\alpha))$.  In Table \ref{nuqam_table} we give examples of some such numerically-optimized NUQAM constellations, and in Fig.\ \ref{nuqam_scatter} we plot an example of a 1024-NUQAM constellation.

Of course, one could maximize the cutoff rate of $\mathcal{X}$ by performing steepest descent on every coordinate of every signal point.  However, when $|\mathcal{X}| = 2^{12}$, for example, this approach would require optimization over $2\cdot 2^{12} = 8192$ real parameters, whereas optimizing with respect to non-uniformity parameters requires only $2^{12/2 - 1} = 32$ real parameters.  In the next subsection we shall see how to further improve the cutoff rate by the addition of a single real parameter to the search space, namely one governing rotation.

\subsection{Optimal Rotation Angles}\label{opt_rot_2D}

Let us study how to further improve the cutoff rate of constellations in $\R^2$ by rotating them.  While this offers large benefits in terms of rate, we note that rotated constellations are necessarily more difficult to decode, since one cannot decode the coordinates independently.  We optimize the rotation angle of a constellation $\mathcal{X}$ by considering the objective function
\begin{equation}\label{2dobj}
R({\bf Q}_2(t)\mathcal{X}),\ \ {\bf Q}_2(t) = 
\begin{bmatrix*}[r]
\cos(t) & \sin(t) \\
-\sin(t) & \cos(t)
\end{bmatrix*}
\end{equation}
and performing a brute force search over the interval $t\in [0,\pi/2]$.
\begin{figure}[h!]
\centering
\includegraphics[width = .45\textwidth]{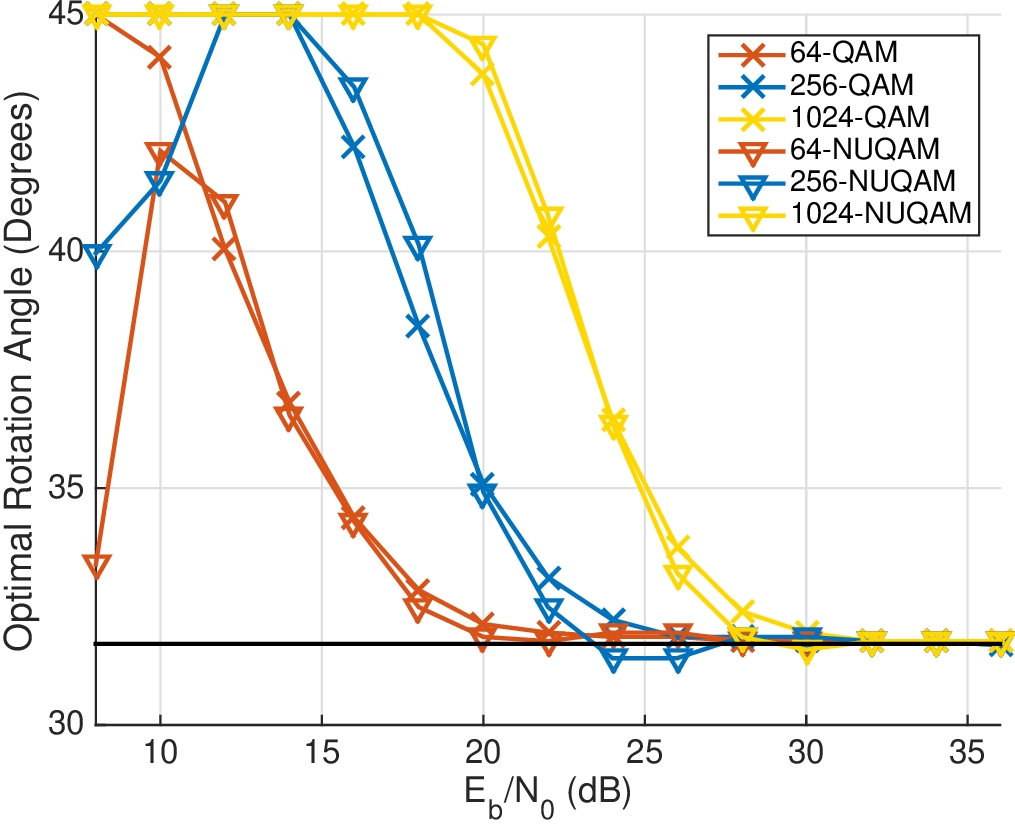}\hfill
\includegraphics[width = .45\textwidth]{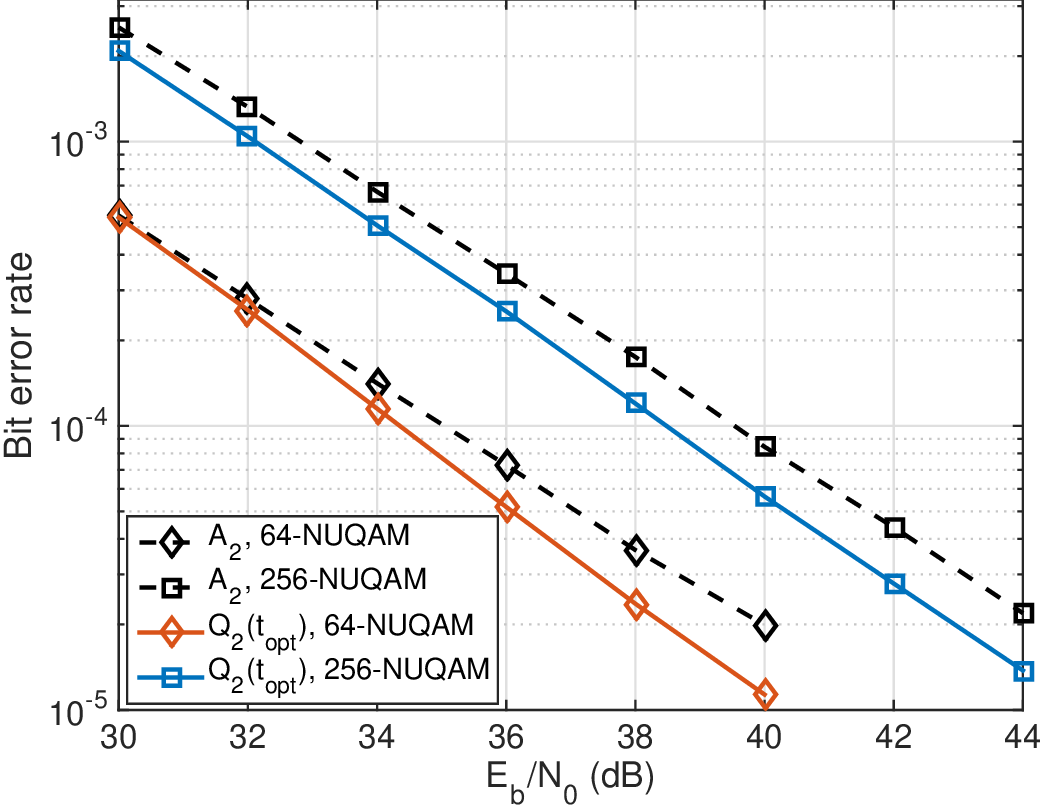}
\caption{\textcolor{black}{On the left, we plot} optimal rotation angles for $M$-QAM and $M$-NUQAM constellations in $\R^2$, for $M = 64$, $256$, and $1024$.  The horizontal black line represents the angle corresponding to the optimal two-dimensional algebraic rotation matrix ${\bf A}_2$ \cite{Viterbo_rotations}.  \textcolor{black}{On the right, we plot the bit error rate of the $64$- and $256$-NUQAM constellations, rotated by the optimal rotation matrix ${\bf Q}_2(t_{\text{opt}})$ as well as the algebraic rotation ${\bf A}_2$.} }\label{optimal_angles_2D}
\end{figure}

To motivate the need for constellation- and SNR-dependent rotations, we plot in Fig.\ \ref{optimal_angles_2D} the rotation angle $t_{\text{opt}}\in [0,\pi/2]$ which optimizes the objective function in (\ref{2dobj}) as a function of $E_b/N_0$.  The NUQAM constellations used were optimized for each value of $E_b/N_0$ and then optimal rotation angles were considered.  It is clear from Fig.\ \ref{optimal_angles_2D} that the optimal rotation angle depends on the modulation order, the SNR, and the uniformity of the constellation.  

Also in Fig.\ \ref{optimal_angles_2D} we plot the \textcolor{black}{bit} error rate for the $64$- \textcolor{black}{and $256$}-NUQAM constellations, rotated by the optimal matrix ${\bf Q}_2(t_{\text{opt}})$ as well as the optimal algebraic rotation ${\bf A}_2$ \cite{Viterbo_rotations}.  \textcolor{black}{Bit strings of length at least $10^8$ were modulated to the constellation points at every value of $E_b/N_0$, according to a Gray labeling.  We see that the gain obtained by rotating on a per-SNR basis can be as high as $1.5$ dB at high values of $E_b/N_0$ for both NUQAM constellations.  While the operating range for the SNR is quite high, we are ultimately interested in comparison with the algebraic lattices of \cite{OBV} which are designed for the very high SNR regime.  Thus the main point is not the magnitude of the gains but rather that there are any gains at all, especially in the very high regime for which algebraic rotations are often considered optimal.}

These plots provide motivation for finding SNR- and constellation-dependent rotation matrices for constellations in $\R^2$ as well as in larger ambient spaces.   The algebraic and number-theoretic methods of \cite{OV,Viterbo_rotations} of constructing good rotated $\Z^n$-lattices necessarily fall somewhat short of this goal since NUQAM constellations are not subsets of lattices, and furthermore algebraic lattices are not SNR-dependent.


\begin{figure}[h!]
\centering
\includegraphics[width = .45\textwidth]{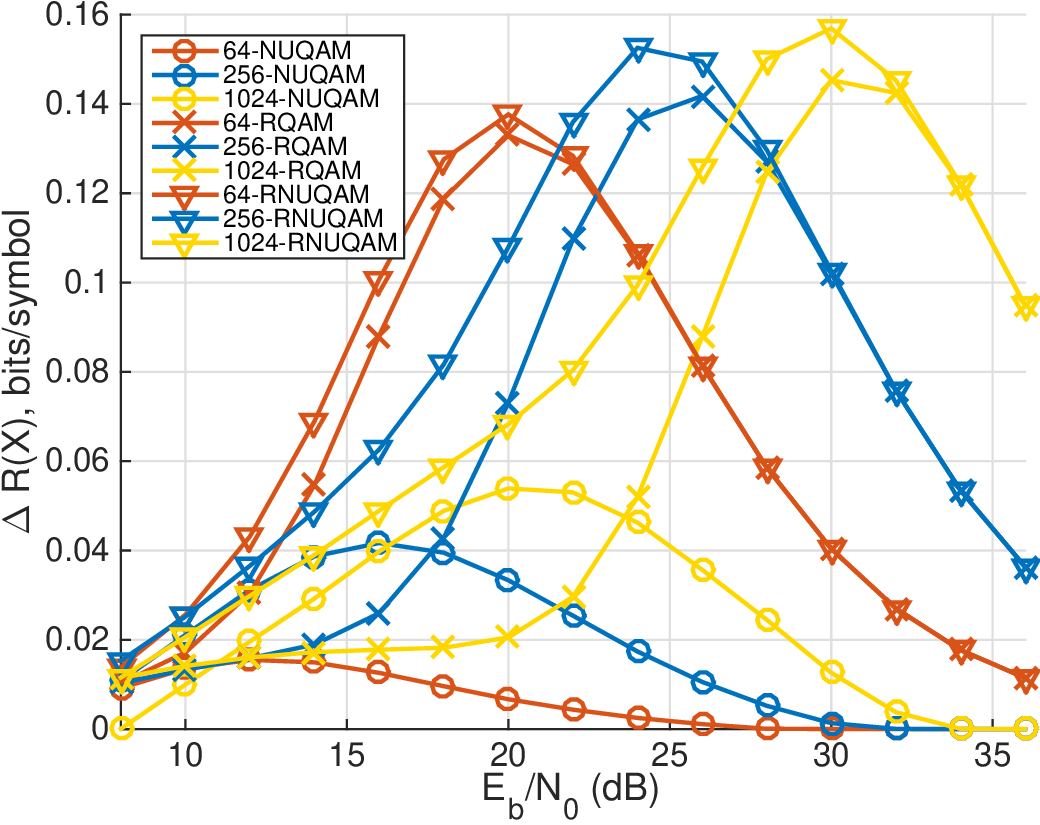}\hfill 
\includegraphics[width = .45\textwidth]{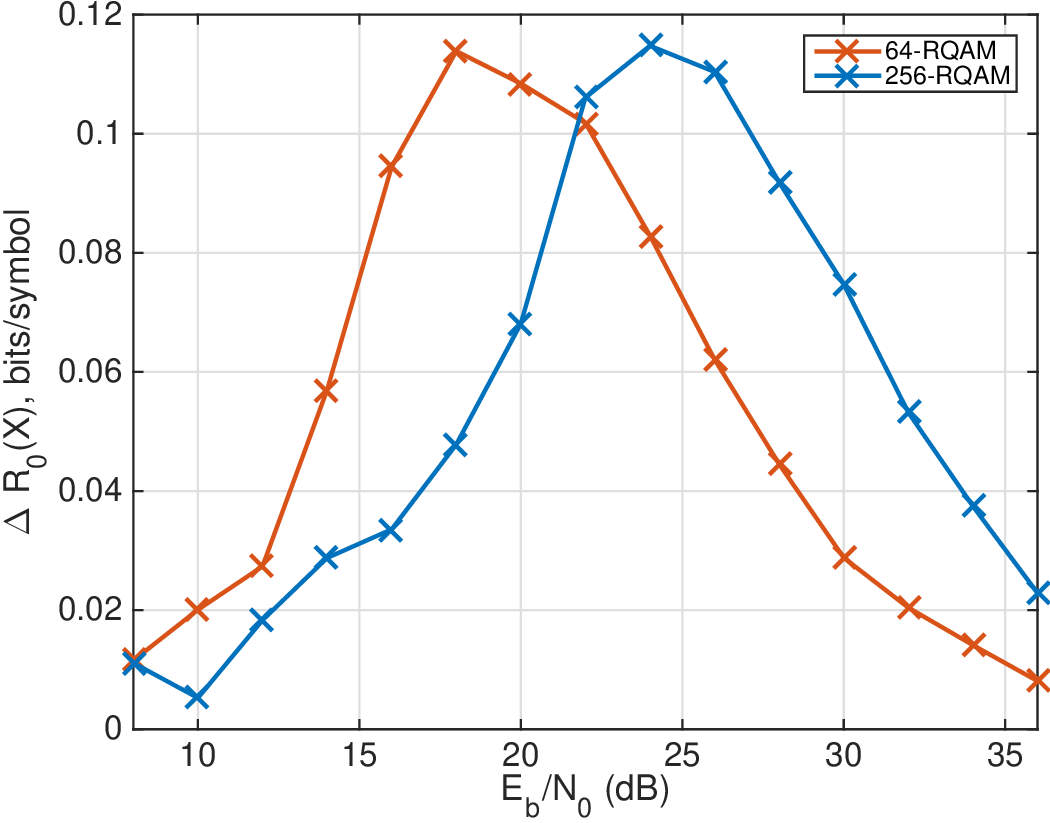}
\caption{\textcolor{black}{On the left, we plot the} improvement in $R(\mathcal{X})$ for non-uniform (NUQAM), rotated (RQAM), and rotated non-uniform (RNUQAM) constellations over their unrotated, uniform counterparts.  \textcolor{black}{On the right, we plot the experimentally measured increase in $R_0(\mathcal{X})$ for $64$-RQAM and $256$-RQAM constellations.}}\label{R_improve_2D}
\end{figure}

In Fig.\ \ref{R_improve_2D} we plot the overall increases
\begin{equation}
\textcolor{black}{\Delta R(\mathcal{X}) = R(\mathcal{X}) - R(\mathcal{X}_{\text{QAM}}),\quad \Delta R_0(\mathcal{X}) = R_0(\mathcal{X}) - R_0(\mathcal{X}_{\text{QAM}})}
\end{equation}
\textcolor{black}{in cutoff rate for various constellations $\mathcal{X}$, over their uniform, unrotated counterparts of the same modulation order, denoted by $\mathcal{X}_{\text{QAM}}$.  In measuring the increase in $R(\mathcal{X})$,} we used non-uniform (NUQAM), rotated (RQAM), and rotated non-uniform (RNUQAM) constellations, for $M = 64$, $256$, and $1024$, over their uniform, unrotated counterparts.  For the NUQAM and RQAM constellations, the non-uniformity and rotation parameters were numerically optimized by performing steepest descent on the cutoff rate.  The RNUQAM constellations were first optimized with respect to the non-uniformity parameters, and then with respect to rotation.  Notice that non-uniformity and rotation can improve the cutoff rate of the constellation at different values of $E_b/N_0$.  For example at $E_b/N_0 = 20$ dB, $1024$-QAM benefits much more from introducing non-uniformity than from rotation.  Lastly we note that the benefit of rotation is most pronounced in the finite SNR regime.

\textcolor{black}{The quantity $\Delta R_0(\mathcal{X})$ was computed for $64$-RQAM and $256$-RQAM by experimentally averaging the value of $R_0(\mathcal{X};{\bf h}) - R_0(\mathcal{X}_{\text{QAM}};{\bf h})$ over $2*10^5$ channels ${\bf h}$.  Here optimal rotations were computed on a per-SNR basis, so that $\mathcal{X}$ was chosen to be the rotated version of $\mathcal{X}_{\text{QAM}}$ which maximizes $R(\mathcal{X})$.  As one can see by comparing the plots, maximizing $R(\mathcal{X})$ results in a significant increase in $R_0(\mathcal{X})$ as well, and thus we include these simulation results to empirically demonstrate the strong correlation between these two objective functions.  However, due to the difficulty of computing $R_0(\mathcal{X})$ even for low dimensions, we will forego measuring constellation performance in terms of $R_0(\mathcal{X})$ and only present further simulation results in terms of $R(\mathcal{X})$.}




\section{Local Diversity}

Note that in Fig.\ \ref{optimal_angles_2D} the optimal angle for a number of constellations and SNR values was $t = 45^\circ$, which does \emph{not} offer full diversity (in the sense of \cite{OV} and defined below) in the case of a rotated QAM constellation.  Thus at finite SNR, constellations which are not fully diverse have the potential to outperform fully diverse constellations.  In this section we define a subtler notion of diversity which seeks to explain this phenomenon.

\subsection{Full Diversity and Minimum Product Distance} \label{diversity}

Let $\mathcal{X}\subset\R^n$ be any constellation.  If we let $N_0\rightarrow 0$, that is, consider the high SNR regime, then we could further approximate the cutoff rate (\ref{cutoff_rate}) by
\begin{equation}
R(\mathcal{X})\approx q - \log_2\left[1+2^{-q}S_0\right]
\end{equation}
where
\begin{equation}\label{sum}
S_0 = \sum_{{\bf x} \neq {\bf y}\in\mathcal{X}}\prod_{\substack{i = 1 \\ x_i\neq y_i}}^n\frac{8N_0}{(x_i-y_i)^2}
\end{equation}
To maximize $R(\mathcal{X})$, then, one wishes to minimize $S_0$.  As a first attempt at doing so, one maximizes the number of terms in each of the above products.  For that we consider the \emph{diversity order} $L$, defined by
\begin{equation}\label{div}
L(\mathcal{X}) = \min_{{\bf x} \neq {\bf y}\in\mathcal{X}}\#\{i\ |\ x_i\neq y_i\}
\end{equation}
In other words, $L(\mathcal{X})$ measures the minimum number of coordinates in which two signal points differ.  When $L(\mathcal{X}) = n$ one says that $\mathcal{X}$ has \emph{full diversity}, which is often a desirable property and can be effectively achieved via algebraic methods \cite{OV}.

As a sharper measure of constellation performance, one can then consider the \emph{minimum product distance} $d_p(\mathcal{X})$ defined by
\begin{equation}
d_p(\mathcal{X}) = \min_{{\bf x} \neq {\bf y}\in\mathcal{X}} \prod_{\substack{i = 1 \\ x_i\neq y_i}}^n|x_i-y_i|
\end{equation}
At high SNR design criteria typically reduces to choosing an $\mathcal{X}$ which maximizes $d_p(\mathcal{X})$, for which the number-theoretic constructions of \cite{OV} are effective.

\subsection{Local Cutoff Rate and Local Diversity} \label{localdefns}

The notion of full diversity can be somewhat restrictive.  Indeed, it may not matter that much that all of the coordinates of ${\bf x}$ and ${\bf y}$ are distinct if they are very far from one another in Euclidean distance.  Thus rate and error performance should depend only on local neighborhoods of the constellation and not on its global structure.  We loosen the notion of full diversity in the following way to account for this fact.

Let ${\bf x}\in\mathcal{X}$ and fix a radius $0<r\leq\infty$.  Define the ball of radius $r$ at ${\bf x}$ by
\begin{equation}
B({\bf x},r) = \{{\bf y}\in\mathcal{X}\ |\ ||{\bf x}-{\bf y}||_2\leq r\}.
\end{equation}
The intuition behind the following definition is that the summands which dominate the summation appearing in (\ref{cutoff_rate}) are those for which the points ${\bf x} \neq {\bf y}$ are close in Euclidean distance.

\begin{definition}
Given a radius $0< r \leq \infty$, we define the \emph{local cutoff rate $R(\mathcal{X},r)$} of $\mathcal{X}$ by
\begin{equation}
R(\mathcal{X},r) = q- \log_2(1+2^{-q}S(r))
\end{equation}
where
\begin{equation}
S(r) = \sum_{{\bf x}\in\mathcal{X}}\sum_{\substack{{\bf y}\in B({\bf x},r) \\ {\bf y}\neq {\bf x}}}\prod_{i = 1}^n\frac{1}{1+\frac{(x_i-y_i)^2}{8N_0}}.
\end{equation}
\end{definition}

The accuracy of $R(\mathcal{X},r)$ to predict the value of $R(\mathcal{X})$ depends on the value of $r$, and the $r$ one wishes to consider may in turn depend on the SNR.  For $r= \infty$ we have $R(\mathcal{X},\infty) = R(\mathcal{X})$.  One of the main advantages of considering local cutoff rates is that one can often find explicit provably optimal rotations, as we shall see.  These rotations will not result in fully diverse constellations, but they will be ``locally'' fully diverse in the following sense.
\begin{definition}
Given a radius $0<r\leq \infty$, the \emph{local diversity order of $\mathcal{X}$} is
\begin{equation}
L(\mathcal{X},r) = \min_{{\bf x}\in\mathcal{X}} \min_{\substack{{\bf y}\in B({\bf x},r) \\ {\bf y}\neq {\bf x}}} \#\{i\ |\ x_i\neq y_i\}.
\end{equation}
If $L(\mathcal{X},r) = n$ for some radius $r$ then we shall say that $\mathcal{X}$ has \emph{local full diversity}.  If $L(\mathcal{X},\infty) = n$ we shall say that $\mathcal{X}$ has \emph{global full diversity}.
\end{definition}

The property of local full diversity of course depends on the particular radius, which we shall always make explicit.  Global full diversity is equivalent to full diversity in the traditional sense of (\ref{div}), but we append the adjective `global' to emphasize the distinction.

For example, consider the constellation
\begin{equation}
\mathcal{X} = \left\{(\pm\sqrt{2},0),(0,\pm\sqrt{2})\right\}\subset\R^2.
\end{equation}
One sees easily that $L(\mathcal{X},2) = 2$ but $L(\mathcal{X},\infty) = 1$.  Hence $\mathcal{X}$ is locally fully diverse but not globally fully diverse.  In fact, $\mathcal{X}$ is the standard $4$-QAM constellation in $\R^2$ rotated by $t=45^\circ$, which the experiments of the previous section show is often an optimal rotation for constellations in $\R^2$.  

Loosely speaking, if the SNR is such that the large majority of errors cause constellation points to be confused only for others within a ball of radius $r$, then the performance of a locally fully diverse constellation should be indistinguishable from that of a globally diverse constellation.  The above definitions reflect the fact that error performance is largely a local invariant of a constellation.

We can compare two locally diverse constellations at high SNR using the following variant of the minimum product distance.
\begin{definition}
Given a radius $0 < r \leq \infty$, the \emph{local minimum product distance of $\mathcal{X}$} is
\begin{equation}
d_p(\mathcal{X},r) = \min_{{\bf x}\in\mathcal{X}}\min_{\substack{y\in B({\bf x},r) \\ {\bf y}\neq {\bf x}}}\prod_{\substack{i = 1 \\ x_i\neq y_i}}^n|x_i - y_i|
\end{equation}
\end{definition}
For $r = \infty$ we have the obvious equality $d_p(\mathcal{X},\infty ) = d_p(\mathcal{X})$.  In Section \ref{n4results} we will show that the local minimum product distance can sometimes predict the relative performance of two constellations when $d_p(\mathcal{X})$ cannot.

\section{The Special Orthogonal Group $SO(n)$}\label{graddescent}

The large gains in cutoff rate offered by rotations at finite SNR for two-dimensional constellations motivate us to find good rotations for higher-dimensional constellations, which are tailored for a given SNR and constellation.  To provide the proper mathematical framework, we first discuss the natural habitat of higher-dimensional rotations.

\subsection{Basic Definitions}
To properly describe the families of rotation matrices we construct, we need to discuss the structure of the \emph{special orthogonal group} $SO(n)$ of all $n$-dimensional rotation matrices.  The group $SO(n)$ of rotations of $\R^n$ is defined by
\begin{equation}
SO(n) = \{{\bf Q}\in GL(n)\ |\ {\bf QQ}^t = {\bf I}_n,\ \det({\bf Q}) = 1\}
\end{equation}
where $GL(n)$ is the group of all invertible real $n\times n$ matrices.  The dimension of $SO(n)$ as a manifold is $(n^2-n)/2$, which can be thought of as the minimum number of parameters required to describe an $n\times n$ rotation matrix.  The special orthogonal group is a \emph{Lie group}, which is both a group and a manifold such that the group operations are continuous with respect to the manifold structure. As a general reference for the theory of Lie groups we recommend \cite{hallie}.  

The \emph{Lie algebra} $\frak{so}(n)$ of $SO(n)$ is the tangent space at the identity matrix ${\bf I}_n\in SO(n)$, and thus is a real vector space of dimension $(n^2-n)/2$.  We have the following convenient explicit description:
\begin{equation}
\frak{so}(n) = \{{\bf A}\in M(n)\ |\ -{\bf A} = {\bf A}^t\}
\end{equation}
where $M(n)$ is the set of all $n\times n$ real matrices.  We pass from the Lie algebra to the Lie group using the exponential map $\exp:\frak{so}(n)\rightarrow SO(n),$ defined by the familiar power series
\begin{equation}
\exp({\bf A}) = {\bf I}_n + {\bf A} + \frac{{\bf A}^2}{2!} + \frac{{\bf A}^3}{3!}+\cdots
\end{equation}
One can verify easily that the exponential series converges for all ${\bf A}$.  There also exists a logarithm map $\log:SO(n)\rightarrow \frak{so}(n)$, defined by a similar power series and satisfying $\exp(\log({\bf A})) = {\bf A}$.  It is a basic fact that the exponential of a skew-symmetric matrix is a rotation matrix, and that the logarithm of a rotation matrix is skew-symmetric.

A \emph{one-parameter subgroup} of $SO(n)$ is the image of a continuous group homomorphism $\R\rightarrow SO(n)$. One can show all one-parameter subgroups are of the form 
\begin{equation}
{\bf Q}(t) = \exp(t{\bf A})
\end{equation}
for some ${\bf A}\in\frak{so}(n)$ \textcolor{black}{(which depends on the particular subgroup)} and $t\in \R$.  The following lemma will prove useful for the particular one-parameter subgroups we will consider.

\begin{lemma}\label{asq}
Suppose that ${\bf A}\in M(n)$ is such that ${\bf A}^2 = -{\bf I}_n$, and let ${\bf Q}(t) = \exp(t{\bf A})$.  Then
\begin{equation}
{\bf Q}(t) = \cos(t){\bf I}_n + \sin(t){\bf A}.
\end{equation}
\end{lemma}
\begin{IEEEproof}
A straightforward computation using the definition of the exponential map yields
\begin{align}
\exp(t{\bf A}) &= {\bf I}_n + t{\bf A} + \frac{t^2{\bf A}^2}{2!} + \frac{t^3{\bf A}^3}{3!} + \cdots \\
&= \left(1 - \frac{t^2}{2!} + \cdots\right){\bf I}_n
+ \left(t - \frac{t^3}{3!} + \cdots\right){\bf A} \\
&= \cos(t){\bf I}_n + \sin(t){\bf A}
\end{align}
as claimed.
\end{IEEEproof}

In the same way that $e^{ti} = \cos(t) + \sin(t)i$ where $i^2 = -1$ traces out the unit circle, the family $\exp(t{\bf A}) = \cos(t){\bf I}_n + \sin(t){\bf A}$ where ${\bf A}\in\frak{so}(n)$, ${\bf A}^2 = -{\bf I}_n$ traces out a circle in the group $SO(n)$.  Thus if one can find a desirable skew-symmetric matrix ${\bf A}$, optimizing over all of $SO(n)$ can be reduced to optimizing over a circle, which can be done by brute force.

While computing the exponential of an arbitrary matrix can be computationally demanding, the above proposition reduces us, in the case where ${\bf A}^2 = -{\bf I}_n$, to computing $\cos(t)$ and $\sin(t)$.  This is useful when computing the optimal $t\in [0,2\pi]$.

\subsection{Gradient Fields and Steepest Descent on $SO(n)$}

Let $f:M(n)\rightarrow \R$ be a \textcolor{black}{differentiable} real-valued function, \textcolor{black}{and suppose that we prescribe coordinates ${\bf q}_{ij}$, $1\leq i,j\leq n$ to the space $M(n)$.}  The gradient of $f$ as a function on $M(n)$ can be identified with the vector field
\begin{equation}
\textcolor{black}{\nabla f:M(n)\rightarrow M(n),\quad \nabla f({\bf Q}) = \left(\frac{\partial f}{\partial q_{ij}}({\bf Q})\right)_{1\leq i,j\leq n},\ {\bf Q} = (q_{ij})_{1\leq i,j\leq n}}
\end{equation}
We wish to understand critical points of $f$ when restricted to $SO(n)$, that is, describe the gradient field $\nabla (f|_{SO(n)})$.  To that end, let us also define the vector field
\begin{equation}
\textcolor{black}{X_f:SO(n)\rightarrow \frak{so}(n),\quad X_f({\bf Q}) = \nabla f({\bf Q}){\bf Q}^t - {\bf Q}\nabla f({\bf Q})^t}
\end{equation}
which is essentially a translated version of $\nabla (f|_{SO(n)})$, but is more convenient to work with since $X_f({\bf Q})$ is skew-symmetric.  The following theorem and its proof can be found in \cite[\S2.4]{edelman}, in which the authors work with a more general structure known as the Stiefel manifold.  In the notation of \cite{edelman}, the Stiefel manifold is equal to the orthogonal group when $p = n$.

\begin{theorem}\label{big_theorem} \cite[\S2.4.4]{edelman}
Let $f:M(n)\rightarrow \R$ be a smooth, real-valued function.  Then for any ${\bf Q}\in SO(n)$, we have
\begin{equation}
\nabla (f|_{SO(n)})({\bf Q}) = X_f({\bf Q}) {\bf Q}
\end{equation}
and thus ${\bf Q}$ is a critical point of $f$ if and only if $X_f({\bf Q}) = 0$.  
\end{theorem}
\begin{IEEEproof} See \cite{edelman}.
\end{IEEEproof}

Let $f:M(n)\rightarrow \R$ be an objective function which we wish to minimize on the subset $SO(n)\subset M(n)$.  To do this, we use the \emph{conjugate gradient} method of \cite[\S3.4]{edelman}, also known as the method of \emph{geodesic flow}.  The conjugate gradient algorithm approximates the shortest path in $SO(n)$ between an initial point ${\bf Q}_0$ and an approximate local minimum ${\bf Q}_N$ of $f$.

The conjugate gradient method presented in \cite{edelman} works on a much more general class of manifolds, but we simplify and summarize the method briefly here for the specific case where the manifold is $SO(n)$.  The conjugate gradient method starts with a rotation matrix ${\bf Q}_0$ and a step size $h$, and recursively updates via the formula
\begin{equation}\label{descent}
{\bf Q}_{k+1} = \exp\left(-hX_f({\bf Q}_k)\right){\bf Q}_k.
\end{equation}
The matrix $X_f({\bf Q}_k)$ is skew-symmetric and should be thought of as an infinitesimal gradient matrix living on the Lie algebra $\frak{so}(n)$.  Thus $\exp\left(-hX_f({\bf Q}_k)\right)$ is an element of $SO(n)$, which restricts this gradient search algorithm to the space of rotation matrices as desired.  For an explanation of why this algorithm finds a local minimum of $f$, see \cite{edelman}.  \textcolor{black}{Note that we do not assume $f$ to be convex, and hence we must content ourselves with finding local rather than global minima of our objective functions.}

Unlike many approaches to optimization over rotation matrices, the above numerical approach avoids the use of explicit parameterizations of rotation matrices.  Such parameterizations become unwieldy and non-canonical as the dimension $n$ grows, hence we prefer the above method which circumvents any attempt at parameterizing the special orthogonal group as is done in \cite{herath}.

\section{Families of $2^k$-Dimensional Rotations}

Given a fixed constellation $\mathcal{X}\subset\R^n$ and some $E_b/N_0$, finding a rotated version of $\mathcal{X}$ which optimizes the cutoff rate (\ref{cutoff_rate}) is equivalent to finding maxima of the function $f:SO(n)\rightarrow \R$ defined by
\begin{equation}\label{objfn}
f({\bf Q}) = R({\bf Q}\mathcal{X})
\end{equation}
In this section we will construct a family of candidates for good rotations for arbitrary constellations $\mathcal{X}\subset \R^{2^k}$ for any $k$.   We first use the numerical conjugate gradient method of (\ref{descent}) to find example critical points, and then demonstrate how to explicitly construct such families of matrices in general, which approximate local maxima of the cutoff rate.

If $\mathcal{X}\subset \R^2$ is an $M$-QAM constellation or any rotated, non-uniform variant thereof, we will refer to $\mathcal{X}^n\subset \R^{2n}$ as a $2n$D $M$-QAM constellation.  Thus by a $4$D $4$-QAM constellation, for example, we mean the set
\begin{equation}\label{4d4qam}
\mathcal{X} = \{(x_1,\ldots,x_4)\in \R^4\ |\ x_i = \pm1\}.
\end{equation}
of size $16$.  An $M$-QAM constellation in $\R^{2n}$ thus contains $M^n$ distinct signal points.

\subsection{Finding Optimal Rotations Using Steepest Descent}

Except for some trivial cases (for example $|\mathcal{X}| = 2$), analytically determining the critical points of $f$ from (\ref{objfn}) seems to be too difficult a task.  On the other hand, the computational simplicity of evaluating the cutoff rate (as opposed to working directly with the CM capacity) allows one to use fairly complicated numerical schemes to find local maxima of $f$ on $SO(n)$.  Thus to obtain good rotations of non-uniform and high-dimensional constellations, we can use the conjugate gradient algorithm of the previous subsection.  

Consider the $4$D $4$-QAM constellation of (\ref{4d4qam}). For $E_b/N_0$ between $8$ and $12$ dB, we ran the conjugate gradient scheme (\ref{descent}) to numerically calculate optimal rotation matrices for this constellation.  For $E_b/N_0 = 10$ dB, for example, the geodesic flow algorithm outputs a rotation matrix ${\bf Q}_N$ such that
\begin{equation}\label{QN}
\log({\bf Q}_{N}) = \begin{bmatrix*}[r]
    0 &   0.73 &   0.73 &   0.73 \\
   -0.73 &   0  &   0.72 &  -0.72 \\
   -0.73 &  -0.72 &  0 &   0.72 \\
   -0.73 &   0.72 &  -0.72 & 0
\end{bmatrix*}
\end{equation}
\textcolor{black}{Here the initial point ${\bf Q}_0$ was chosen to be the matrix ${\bf Q}_0 = \exp({\bf H})$, ${\bf H} = (h_{ij})$, $h_{ij} = 10^{-4}$ for $i>j$, $h_{ij} = -h_{ji}$ for $i<j$, and $h_{ii} = 0$ for all $i = 1,\ldots,n$.  Experiments suggest similar small perturbations of the identity matrix work well as initial points for the conjugate gradient method.}  Similarly structured matrices were obtained for other values of $E_b/N_0$ and other $\mathcal{X}$, namely, that $\log({\bf Q}_N)$ appears to be a constant value times a matrix which depends only on the dimension $n$.  

These numerically-obtained results suggest that to construct such matrices explicitly instead of running the gradient descent scheme (\ref{descent}) for every value of $E_b/N_0$, one should construct a single skew-symmetric matrix ${\bf A}$ and consider exponentiating all of its scalar multiples.

\subsection{Explicit Construction of Rotation Matrices}\label{explicit_construction}

We now demonstrate how to construct families of matrices of the form $\log({\bf Q}_N)$ as in (\ref{QN}), when the ambient space has dimension a power of $2$.  To this end we will construct a one-parameter subgroup ${\bf Q}_{2^k}(t)$ of $SO(2^k)$, which in turn reduces the dimension of the search space from $\dim SO(n) = (n^2-n)/2$ down to one.  This saves significant computation time for dimensions as small as $n = 4$.

First let us recall the definition of the Hadamard matrices ${\bf H}_{2^k}\in M(2^k)$:
\begin{equation}
{\bf H}_1 = [1],\ {\bf H}_{2^k} = \begin{bmatrix*}[r] {\bf H}_{2^{k-1}} & {\bf H}_{2^{k-1}} \\ {\bf H}_{2^{k-1}}  & -{\bf H}_{2^{k-1}}  \end{bmatrix*}
\end{equation}
We now construct skew-symmetric matrices ${\bf A}_{2^k}\in \frak{so}(2^k)$ for $k\geq 1$ recursively in the following way:
\begin{equation}
{\bf B}_1 = [0],\ {\bf B}_{2^k} = \begin{bmatrix*}[r] {\bf B}_{2^{k-1}} & {\bf H}_{2^{k-1}}  \\ -{\bf H}_{2^{k-1}}  & {\bf B}_{2^{k-1}} \end{bmatrix*}
\end{equation}
\begin{equation}
{\bf A}_{2^k} = (2^k-1)^{-1/2}{\bf B}_{2^k}
\end{equation}
The factor of $(2^k-1)^{-1/2}$ is only a convenience that simplifies some expressions in what follows.

Given a fixed $n = 2^k$, we consider the one-parameter family of rotation matrices
\begin{equation}\label{master}
{\bf Q}_{n}(t) = \exp(t{\bf A}_{n})\in SO(n)
\end{equation}
for $t\in \R$.  The following proposition describes the matrices ${\bf Q}_{n}(t)$ explicitly, and relieves us of the computational burden of computing a matrix exponential.

\begin{prop} \label{Qexplicit}
For any dimension $n = 2^k$ we have
\begin{equation}
{\bf Q}_{n}(t) = \cos(t){\bf I}_{n} + \sin(t){\bf A}_{n}.
\end{equation}
\end{prop}
\begin{IEEEproof}
One can show easily by induction that the matrices ${\bf B}_n$ and ${\bf H}_n$ anti-commute, that is, that
\begin{equation}
{\bf B}_{n}{\bf H}_n = -{\bf H}_n{\bf B}_{n}
\end{equation}
for all $n\geq 2$.  Furthermore, it is well-known and easy to show that the Hadamard matrices satisfy ${\bf H}_n^2 = n{\bf I}_n$.  It is now straightforward to show that ${\bf B}_{n}^2 = -(n-1){\bf I}_n$, or in other words that
\begin{equation}
{\bf A}_{n}^2 = -{\bf I}_n.
\end{equation}
The result now follows from Lemma \ref{asq}.
\end{IEEEproof}

The previous proposition also allows us to restrict the parameter space to $t\in[0,\pi/2]$, as every matrix ${\bf Q}_n(t)$ for $t\in(\pi/2,2\pi]$ can be obtained from a matrix ${\bf Q}_n(t)$ for $t\in[0,\pi/2]$ via negation and taking transposes, operations which do not affect constellation performance.  For any constellation $\mathcal{X}\subset \R^{n}$ and a fixed value of $E_b/N_0$, we can now optimize the cutoff rate with respect to rotation by computing
\begin{equation}\label{toot}
t_{\text{opt}} = \underset{t\in[0,\pi/2]}{\argmax}\ R({\bf Q}_n(t)\mathcal{X})
\end{equation}
by simple exhaustive search.  

When $n = 2$ we have
\begin{equation}
{\bf Q}_2(t) = \begin{bmatrix*}[r]
\cos(t) & \sin(t) \\
-\sin(t) & \cos(t)
\end{bmatrix*}
\end{equation}
and hence the family ${\bf Q}_2(t)$ is simply the whole group $SO(2)$.  Finding the optimal rotation simply reduces to exhaustive search over the interval $t\in [0,\pi/2]$, which was exactly the procedure carried out in Section \ref{opt_rot_2D}.

When $n = 4$ one can see easily that ${\bf Q}_4(0.72)$ is approximately the matrix ${\bf Q}_N$ from (\ref{QN}).  The transpose of the matrix ${\bf Q}_4(0.56)$ was considered in conjunction with $\mathcal{X}$ a $4$D $4$-QAM constellation in the DVB-NGH standard \cite{DVBNGH} to minimize the bit error rate at the demapper.  In a sense, one could describe our method as a generalization of the rotation matrices considered by DVB-NGH.



\subsection{Optimal Rotations at Low SNR} \label{lowsnr}

Let us consider an arbitrary $M$-QAM constellation $\mathcal{X}\subset \R^n$ for $n = 2^k$.  Our goal in this subsection is to find good rotations among the family ${\bf Q}_n(t)$ when the SNR is low.  At low SNR, errors are mostly caused by large noise vectors as opposed to deep fades. One make this more precise by choosing ${\bf x} \neq {\bf y}\in \mathcal{X}$ and expanding the reciprocal of the product in the expression (\ref{cutoff_rate}) for the cutoff rate.  We obtain
\begin{align}
\prod_{i = 1}^n\left(1+\frac{(x_i-y_i)^2}{8N_0}\right) &= 1 + \frac{||{\bf x}-{\bf y}||^2_2}{8N_0} + \mathcal{O}\left(\frac{1}{N_0^2}\right)
\end{align}
Thus for large $N_0$, the dominant term in the above is proportional to $||{\bf x}-{\bf y}||^2_2$, and the dominant terms in the sum (\ref{cutoff_rate}) are those for which the Euclidean distance is smallest.  Hence if we transmit ${\bf x}$ and decode $\hat{{\bf x}}\neq {\bf x}$, it is overwhelmingly likely that $||{\bf x}-\hat{{\bf x}}||_2 = 2$, which is the minimal Euclidean distance between any two QAM constellation points.  

The previous paragraph suggests that it suffices to consider the approximation
\begin{equation}
R(\mathcal{X})\approx R(\mathcal{X},2)
\end{equation}
of the cutoff rate by the local cutoff rate.  Moreover, this approximation is convenient because it allows us to compute the rotation parameter $t$ optimizing $R({\bf Q}_n(t)\mathcal{X},2)$ explicitly.

\begin{prop}\label{opt_low_SNR}
Let $\mathcal{X}\subset \R^n$ be a QAM constellation and let $n = 2^k$.  The function $f:[0,\pi/2]\rightarrow \R$ defined by
\begin{equation}
f(t) = R({\bf Q}_n(t)\mathcal{X},2)
\end{equation}
has a single maximum at $t = \arccos(1/\sqrt{n})$.
\end{prop}
\begin{IEEEproof}
Let ${\bf Q}\in SO(n)$ be any rotation matrix and consider the summation inside $R({\bf Q}\mathcal{X},r)$ defined by
\begin{equation}
S(r) = \sum_{{\bf x}\in {\bf Q}\mathcal{X}}\sum_{\substack{{\bf y}\in B({\bf x},r) \\ {\bf y}\neq {\bf x}}}\prod_{i = 1}^n\frac{1}{1 + \frac{(x_i - y_i)^2}{8N_0}}.
\end{equation}
For a fixed ${\bf x}\in\mathcal{X}$, we have ${\bf y}\in B(x,2)$ if and only if ${\bf x}-{\bf y} = \pm 2e$ for some standard basis vector ${\bf e}\in\R^n$.  Note that if ${\bf e} = {\bf e}_j$ where $j = 1,\ldots,n$ then ${\bf Qx} - {\bf Qy}$ is the $j^{th}$ column of ${\bf Q}$, whose $i^{th}$ coordinate is $q_{ij}$.  Since ${\bf Q}$ is orthogonal and hence distance-preserving, it follows that ${\bf Q}B({\bf x},r) = B({\bf Qx},r)$ and hence
\begin{equation}
S(r) = \sum_{{\bf x}\in\mathcal{X}}\sum_{\substack{{\bf y}\in B({\bf x},r) \\ {\bf y}\neq {\bf x}}} \prod_{i = 1}^n \frac{1}{1 + \frac{q_{ij}^2}{2N_0}}
\end{equation}
By the symmetry of the constellation, one can see that for all $j$ the product $\prod_{i = 1}^n\frac{1}{1 + q_{ij}^2/2N_0}$ appears in the above summand the same number of times.  Now setting ${\bf Q} = {\bf Q}_n(t)$ we obtain for some positive integer $c$ the expression
\begin{equation}
S(r) = c\sum_{j = 1}^n\prod_{i = 1}^n\frac{1}{1 + \frac{q_{ij}^2}{2N_0}}
= \frac{cn}{g(t)}
\end{equation}
where $g(t)$ is the function
\begin{equation}
g(t) = \left(1+\frac{\cos^2(t)}{2N_0}\right)\left(1+\frac{\sin^2(t)}{(n-1)2N_0}\right)^{n-1}.
\end{equation}
We see that maximizing $R({\bf Q}_n(t)\mathcal{X},2)$ with respect to $t$ is equivalent to maximizing $g(t)$.  One can compute the derivative of $g(t)$ explicitly to see that it has a single maximum on the interval $t\in [0,\pi/2]$ at $t = \arccos(1/\sqrt{n})$ which completes the proof.
\end{IEEEproof}

We should expect from Proposition \ref{opt_low_SNR} that $t = \arccos(1/\sqrt{n})$ is  a decent approximation of the optimal $t$ at low SNR, whenever $R(\mathcal{X},2)$ is a decent approximation of $R(\mathcal{X})$.  This is confirmed via experiment for $n = 2$ in Fig.\ \ref{optimal_angles_2D} and as we shall see in the next section, for $n = 4$ in Fig.\ \ref{opt_rot_angles_4D} and for $n = 8$ in Fig.\ \ref{opt_rot_angles_8D}.

\subsection{Diversity of the Constellations ${\bf Q}_n(t)\mathcal{X}$}

Here we present a short proof of the fact that if $\mathcal{X}\subset \R^n$ is a QAM constellation, then the rotated constellation ${\bf Q}_n(t)\mathcal{X}$ is locally fully diverse for $r = 2$ and all $t\in (0,\pi/2)$.  Furthermore, we show by example that these rotated constellations lack full diversity.  We will show in the next section that our locally fully diverse constellations can outperform the globally diverse constellations defined by number-theoretic rotations of \cite{OV,Viterbo_rotations}, even though our rotated constellations are not globally fully diverse.  

\begin{prop}\label{local_diversity}
Let $\mathcal{X}\subset\R^n$ be a QAM constellation and let ${\bf Q}=(q_{ij})_{1\leq i,j\leq n} \in SO(n)$ be any rotation matrix.  Then ${\bf Q}\mathcal{X}$ satisfies $L({\bf Q}\mathcal{X},2) = n$ if and only if $q_{ij} \neq 0$ for all $i,j$.
\end{prop}
\begin{IEEEproof}
If ${\bf x}\in\mathcal{X}$ and ${\bf y}\in B(x,2)$ is not equal to ${\bf x}$, then ${\bf x}-{\bf y}=2{\bf e}$ for some standard basis vector ${\bf e}$.  Conversely, every standard basis vector of $\R^n$ can be expressed as ${\bf e} = \frac{1}{2}({\bf x}-{\bf y})$ for some ${\bf x}\in\mathcal{X}$ and ${\bf y}\in B(x,2)$.  Hence ${\bf Q}\mathcal{X}$ being locally fully diverse for radius $r = 2$ is equivalent to no entry of ${\bf Qe}$ being zero for any standard basis vector ${\bf e}$.  But this is exactly the statement that $q_{ij} \neq 0$ for all $i,j$.
\end{IEEEproof}

For any choice of $t\in (0,\pi/2)$, it is now clear from Propositions \ref{Qexplicit} and \ref{local_diversity} that for $\mathcal{X}\subset\R^n$ a QAM constellation with $n = 2^k$, the rotated version will satisfy
\begin{equation}
L({\bf Q}_n(t)\mathcal{X},2) = n
\end{equation}
that is, the rotated constellation will be locally fully diverse.  In fact, Proposition \ref{local_diversity} essentially says that it is very easy to achieve local full diversity; simply pick a rotation matrix without any zero entries, and such matrices are dense in $SO(n)$.  Thus the property of local full diversity is not particular to the family ${\bf Q}_n(t)$; rather, among all rotations which offer local full diversity, we study these because they were so frequently the outcome of the numerical algorithm (\ref{descent}). 

However, the rotations ${\bf Q}_n(t)$ do \emph{not} offer full diversity when $n > 2$.  Let us take, for example, $n = 4$, and consider ${\bf z} = {\bf x}-{\bf y}$ where ${\bf x} \neq {\bf y}\in\mathcal{X}$.  Then
\begin{equation}
{\bf Q}_n(t)\begin{bmatrix} z_1 \\ z_2 \\ z_3 \\ z_4 \end{bmatrix} = \begin{bmatrix}
az_1 + b(z_2 + z_3 + z_4) \\
az_2 + b(-z_1 + z_3 - z_4) \\
az_3 + b(-z_1 - z_2 + z_4) \\
az_4 + b(-z_1 + z_2 - z_3)
\end{bmatrix}
\end{equation}
where $a = \cos(t)$ and $b = \sin(t)/\sqrt{3}$.  Now notice that for $z_1 = z_2 = 0$ and $z_3 = -z_4$ we have $({\bf Q}_n(t){\bf z})_1 = 0$.  It is easy to check that the points ${\bf x} = (1, 1, 1, -1)^t$ and ${\bf y} = (1,1,-1,1)^t$ belonging to the standard $4$D $4$-QAM constellation will satisfy $({\bf Q}_n(t){\bf z})_1 = 0$, resulting in a rotated constellation that is not fully diverse.

\section{Simulation Results for $n = 4$ and $n = 8$} 

In this section we confirm the benefits of the family ${\bf Q}_n(t)$ of rotation matrices for $n = 4,8$.  Our basis for comparison is always the optimal algebraic rotation of the same dimension, as found in \cite{Viterbo_rotations}.  With regard to an algebraic rotation, ``optimal'' means with respect to the normalized minimum product distance of the rotated constellation.  Maximum-likelihood decoding was used for all error rate simulations.

\subsection{Rotated Constellations in $\R^4$}\label{n4results}
Let us further examine the family (\ref{master}) for the simplest non-trivial example, that of rotated constellations in $\R^4$.  It follows from Proposition \ref{Qexplicit} that
\begin{equation}
{\bf Q}_4(t) = \cos(t){\bf I}_4 + \frac{\sin(t)}{\textcolor{black}{\sqrt{3}}}\begin{bmatrix*}[r]
0 & 1 & 1 & 1 \\
-1 & 0 & 1 & -1 \\
-1 & -1 & 0 & 1 \\
-1 & 1 & -1 & 0
\end{bmatrix*}
\end{equation}
We note that this one-parameter subgroup of $SO(n)$ was implicitly considered by the DVB-NGH standard \cite[\S6.4]{DVBNGH} when rotating $4$D $4$-QAM constellations.  The DVB consortium specifies the particular rotation parameter $t$ by instead specifying the parameter $\sin(t)$, but this difference is immaterial.

\begin{figure}[h!]
\centering
\includegraphics[width = .45\textwidth]{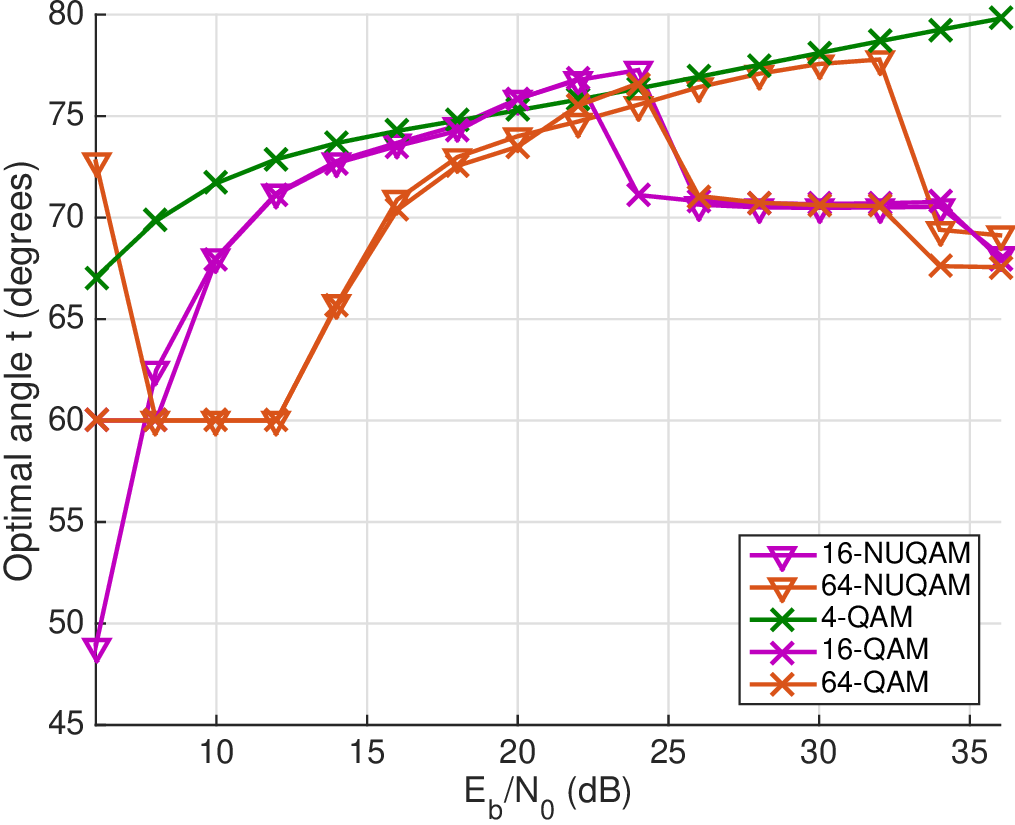}\hfill
\includegraphics[width = .45\textwidth]{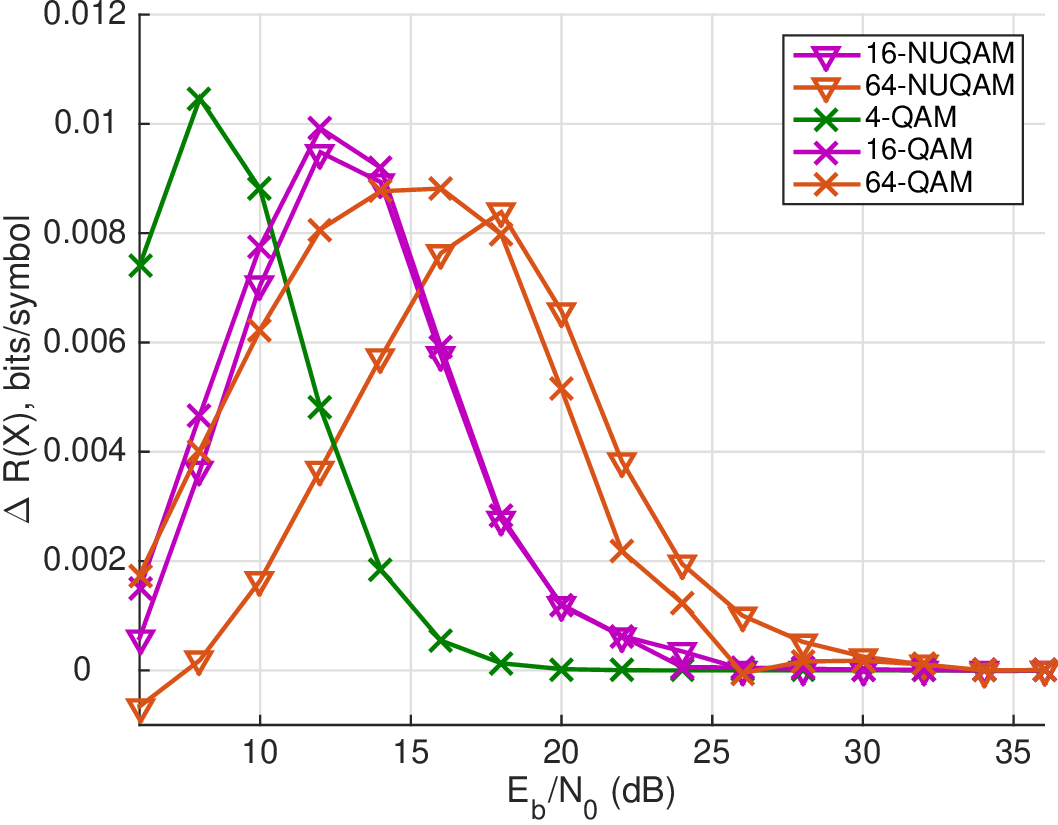}
\caption{On the left, the optimal rotation parameter $t_{\text{opt}}$ as in (\ref{toot}) as a function of $E_b/N_0$, for 4D $M$-QAM and $M$-NUQAM, $M = 4, 16, 64$.  On the right, the improvement in $R(\mathcal{X})$ for 4D $M$-QAM and $M$-NUQAM constellations for $M = 4$, $16$, and $64$ when using the rotation matrices ${\bf Q}_4(t_{\text{opt}})$ as opposed to the Kr\"uskemper rotation, which is the optimal algebraic rotation in $\R^4$ \cite{Viterbo_rotations}.} \label{opt_rot_angles_4D}
\end{figure}

For $4$D $M$-QAM and $M$-NUQAM and $M = 4,16,64$, we computed the optimal rotation matrix ${\bf Q}_4(t_{\text{opt}})$ by exhaustive search over the interval $t\in [0,\pi/2]$ as in (\ref{toot}) and plotted $t_{\text{opt}}$ in degrees as a function of $E_b/N_0$ in Fig.\ \ref{opt_rot_angles_4D}.  When $M = 64$, note the range of relatively low $E_b/N_0$ values for which $t_{\text{opt}} = \arccos(1/\sqrt{4}) = 60^\circ$, as predicted by Proposition \ref{opt_low_SNR}.  The NUQAM constellations in this simulation were first optimized on a per-SNR basis, and then optimal rotations were computed.

To compare our method with the optimal number-theoretic construction, we also plot in Fig.\ \ref{opt_rot_angles_4D} the improvement 
\begin{equation}
\Delta R(\mathcal{X}) = R({\bf Q}_4(t_{\text{opt}})\mathcal{X}) - R({\bf K}_4\mathcal{X})
\end{equation}
of our method over the algebraic method.  Here ${\bf K}_4$ is the Kr\"uskemper rotation, which is the optimal algebraic rotation of $\R^4$ and whose generator matrix is available at \cite{Viterbo_rotations}.  Our rotations ${\bf Q}_4(t_{\text{opt}})$ offer modest but consistent improvement in cutoff rate over the Kr\"uskemper rotation at almost every value of $E_b/N_0$ and every $\mathcal{X}\subset\R^4$ we investigated.  \textcolor{black}{Despite the small improvement in cutoff rate at high SNR, this is the regime where the rotations ${\bf Q}_4(t_{\text{opt}})$ offer the highest error performance gains over ${\bf K}_4$, as we see below.}

\begin{figure}[h!]
\centering
\includegraphics[width = .45\textwidth]{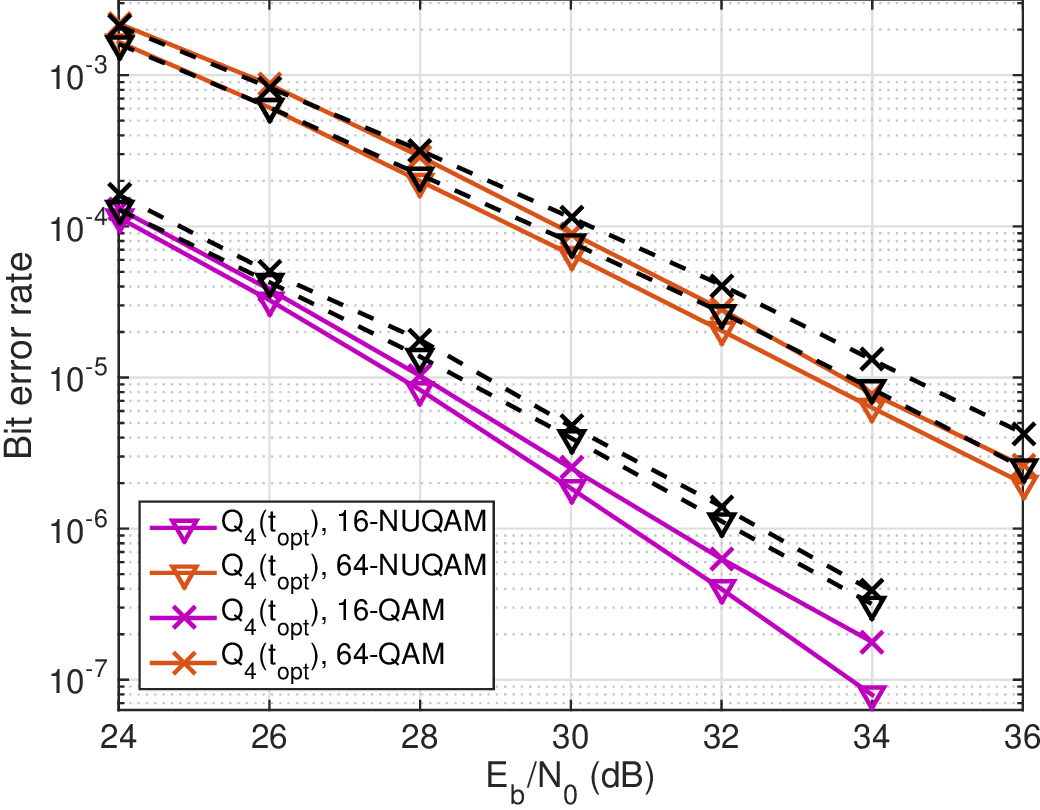}
\caption{Bit error rates for the family ${\bf Q}_4(t_{\text{opt}})$ for $M$-QAM and $M$-NUQAM in $\R^4$, for $M = 16,64$.  The Kr\"uskemper rotation, which is the optimal algebraic rotation in $\R^4$, served as the basis for comparison and is plotted with the analogous dashed black curves. \textcolor{black}{Simulations were performed by modulating approximately $10^9$ bits at each value of $E_b/N_0$.}}\label{SER_4D} 
\end{figure}

We plot in Fig.\ \ref{SER_4D} bit error rates of the family ${\bf Q}_4(t_{\text{opt}})$ for $M$-QAM and $M$-NUQAM, for $M = 16,64$.  Bit strings were assigned according to the Gray labeling of the underlying QAM symbols.  \textcolor{black}{The rotations ${\bf Q}_4(t_{\text{opt}})$ outperform the Kr\"uskemper rotation for both uniform and non-uniform constellations at all values of $E_b/N_0$.  When $M = 16$, we see an increase in performance of up to $2$ dB for non-uniform constellations and about $1.5$ dB for uniform constellations.  When $M = 64$, the increase is approximately $1$ dB for uniform constellations, with a smaller increase in performance for non-uniform constellations.}  

\textcolor{black}{While the operating SNR is very high, we have provided plots in this regime to study asymptotic behavior (i.e.\ diversity).  This allows for fair comparison with the algebraic rotations of \cite{OBV} which are specifically tailored for the asymptotic regime.  The crucial point is not the magnitude of the gains over the fully-diverse algebraic rotations, but rather that there are any gains at all in the absence of full diversity even in the asymptotic regime.  Thus Fig.\ \ref{SER_4D} underscores the original theoretical observation motivating local diversity, namely that full diversity is too restrictive a notion even in the very high SNR regime.  In fact, one can show easily that $L({\bf Q}_4(t_{\text{opt}})\mathcal{X},\infty) = 3$ for all $t_{\text{opt}}$ in the given $E_b/N_0$ range, that is that the rotations outperform the Kr\"uskemper rotation even though their global diversity is in some sense deficient.}

\begin{figure}[h!]
\centering
\includegraphics[width = .45\textwidth]{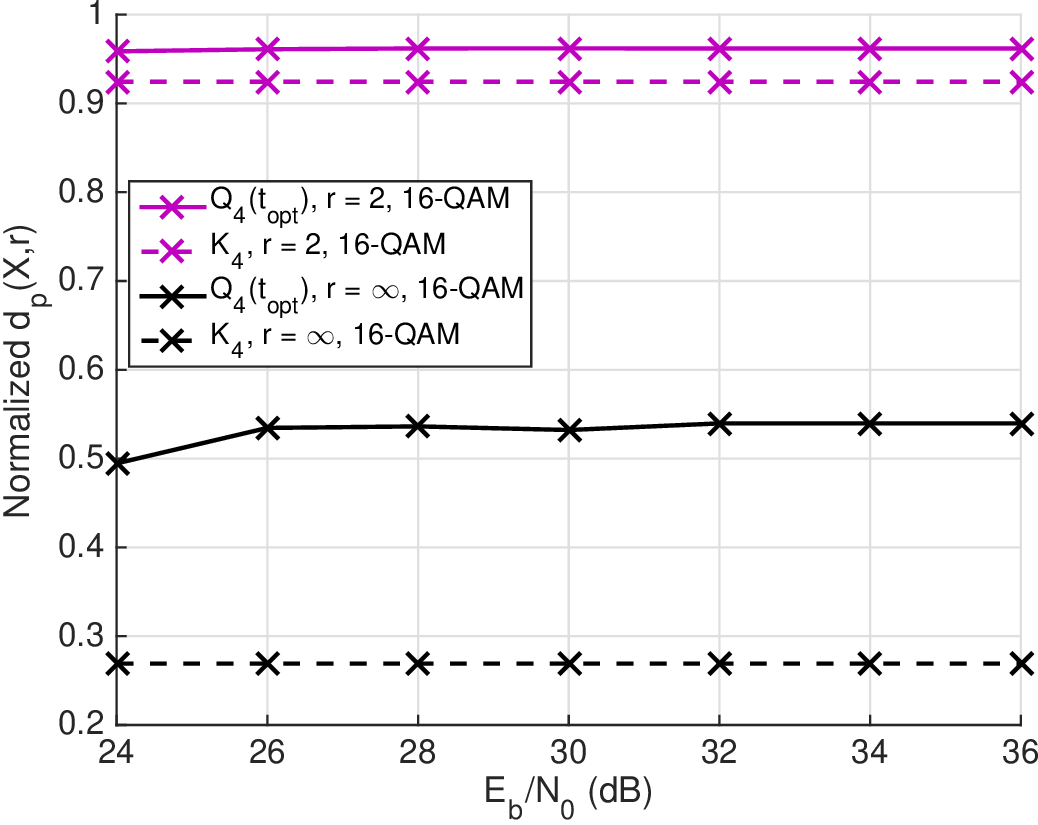} \hfill
\includegraphics[width = .45\textwidth]{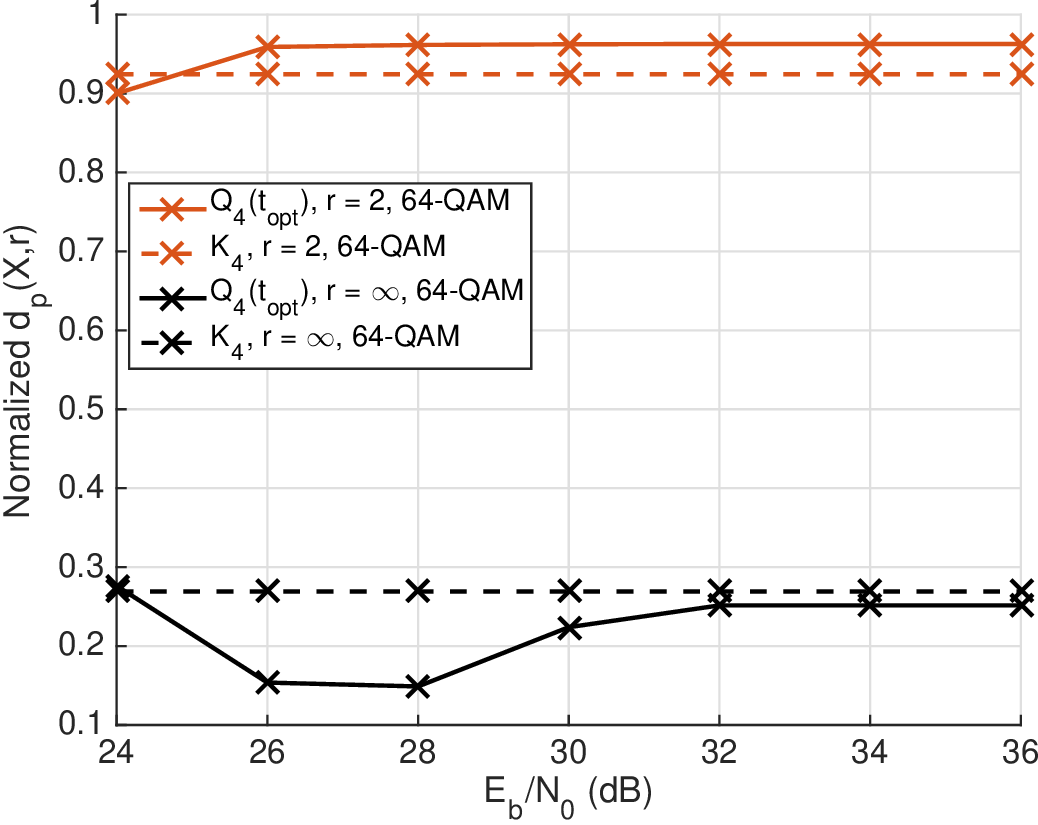}
\caption{Local (with $r = 2$) and global minimum product distances $d_p(\mathcal{X},r)$ for the constellations rotated algebraically with the matrix ${\bf K}_4$, and with the family ${\bf Q}_4(t_{\text{opt}})$.  In the left figure are the results for $4$D $16$-QAM, and in the right figure are the results for $4$D $64$-QAM.}\label{mpds4D}
\end{figure}

Finally in Fig.\ \ref{mpds4D} we plot the various local and global minimum product distances $d_p({\bf Q}\mathcal{X},r)^{1/4}$ for $r = 2,\infty$ as defined in Section \ref{localdefns}, for both ${\bf Q} = {\bf K}_4$ and ${\bf Q} = {\bf Q}_4(t_{\text{opt}})$, and $\mathcal{X} = $ $4$D $M$-QAM for $M = 4,16$.  We see that for $M = 16$, both minimum product distances correctly predict that ${\bf Q}_4(t_{\text{opt}})$ outperforms ${\bf K}_4$.  However when $M = 64$, the local minimum product distance correctly predicts that the family ${\bf Q}_4(t_{\text{opt}})$ outperforms the rotation ${\bf K}_4$ for almost all values of $E_b/N_0$, whereas the global minimum product distance effectively makes a mistake for almost all values of $E_b/N_0$.  Hence the local minimum product distance seems a more appropriate measure of the performance of these constellations.  

\subsection{Rotated Constellations in $\R^8$}\label{n8results}

\begin{figure}[h!]
\centering
\includegraphics[width = .45\textwidth]{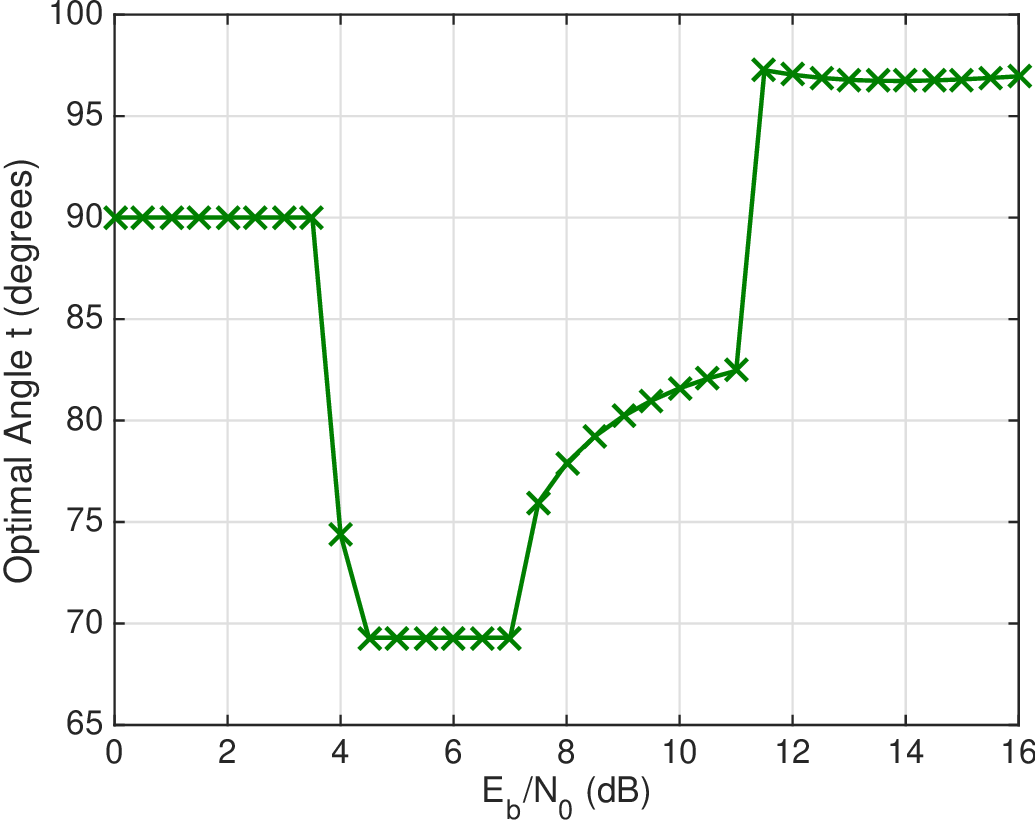}\hfill
\includegraphics[width = .45\textwidth]{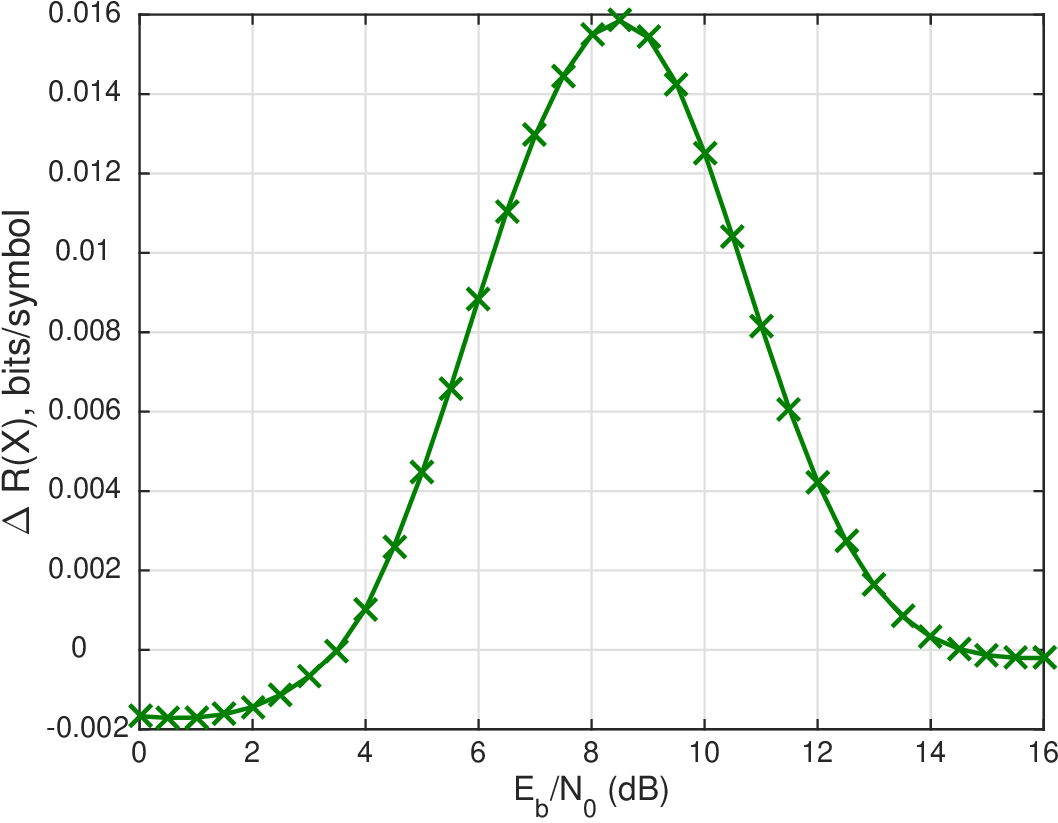}
\caption{On the left, optimal rotation parameter $t_{\text{opt}}$ as in (\ref{toot}) as a function of $E_b/N_0$, for 8D $4$-QAM.  On the right, improvement in $R(\mathcal{X})$ for 8D $4$-QAM when using ${\bf Q}_8(t_{\text{opt}})$ as opposed to the algebraic rotation ${\bf C}_8$ of \cite{Viterbo_rotations}, which is the best-known algebraic rotation of $\R^8$.}\label{opt_rot_angles_8D}
\end{figure}

In this subsection we study rotations of the $8$D $4$-QAM constellation $\mathcal{X}\subset\R^8$.  In Fig.\ \ref{opt_rot_angles_8D} we plot the value of $t_{\text{opt}}$ as a function of $E_b/N_0$.  As predicted by Proposition \ref{opt_low_SNR}, we obtain $t_{\text{opt}} = \arccos(1/\sqrt{8}) = 69.2952^\circ$ for several low values of $E_b/N_0$, approximately in the range $4$-$7$ dB.  We also plot in Fig.\ \ref{opt_rot_angles_8D} the improvement
\begin{equation}
\Delta R(\mathcal{X}) = R({\bf Q}_8(t_{\text{opt}})\mathcal{X}) - R({\bf C}_8\mathcal{X})
\end{equation}
where ${\bf C}_8$ is the optimal algebraic rotation in $\R^8$ \cite{Viterbo_rotations}. We see from the plot that in the range of $E_b/N_0 = 4$-$14$ dB, the family ${\bf Q}_8(t_{\text{opt}})$ shows modest but consistent improvement over ${\bf C}_8$ in terms of cutoff rate, while the algebraic rotation has better cutoff rate outside of this region.

\begin{figure}[h!]
\centering
\includegraphics[width = .45\textwidth]{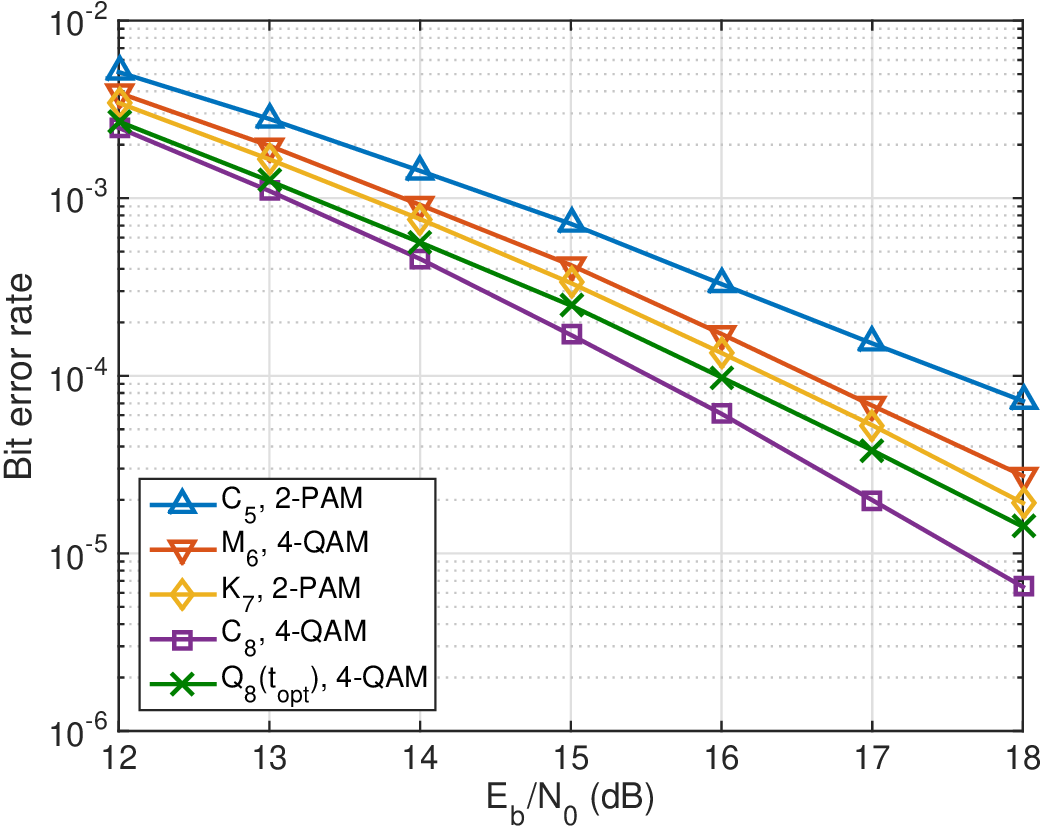}
\caption{Bit error rates for the family ${\bf Q}_8(t_{\text{opt}})$ for $4$-QAM in $\R^8$.  The optimal algebraic rotations of $\R^n$ for $n = 5,6,7,8$, denoted respectively by ${\bf C}_5,{\bf M}_6,{\bf K}_7,{\bf C}_8$, served as the basis for comparison \cite{Viterbo_rotations}.  Simulations were performed by modulating approximately $10^8$ bits at each value of $E_b/N_0$ according to the Gray labeling on the underlying  per-coordinate constellation.  A $2$-PAM (pulse amplitude modulation) constellation was used per real axis for the odd-dimensional constellations, to ensure fair comparison across dimensions.}\label{SER_8D}
\end{figure}

In Fig.\ \ref{SER_8D} we plot the bit error rate of the rotations ${\bf Q}_8(t_{\text{opt}})$ as well as those of the optimal algebraic rotations of $\R^n$ for $n = 5,6,7,8$, denoted by ${\bf C}_5,{\bf M}_6,{\bf K}_7,{\bf C}_8$, respectively \cite{Viterbo_rotations}.  In the low SNR regime the performance of the two schemes is identical. Beyond $E_b/N_0>12$ dB (corresponding to a bit error rate of approximately $10^{-3}$) the algebraic rotation outperforms the rotations ${\bf Q}_8(t_{\text{opt}})$.

While the rotations ${\bf Q}_8(t_{\text{opt}})$ have inferior error rate performance compared to ${\bf C}_8$ at high SNR, it can be shown that the global diversity of our scheme is only $L({\bf Q}_8(t_{\text{opt}})\mathcal{X},\infty) = 5$ whereas the fully-diverse algebraic rotations of $\R^n$ achieve global diversity $n$.  Thus the error plots show that ${\bf Q}_8(t_{\text{opt}})$ outperforms rotations whose global diversity is up to $n = 7$, lending credence to the original theoretical notion that local diversity should be a superior predictor of performance.

Lastly, we mention that neither local nor global minimum product distances appeared to predict the relative error performance, hence we omit plots of these quantities.

\section{Conclusions and Future Work}\label{conclusions}

We have constructed a family of rotation matrices ${\bf Q}_{2^k}(t)\in SO(2^k)$ for every $k$ with the goal of maximizing the \textcolor{black}{cutoff rate} of arbitrary constellations in $\R^n$.  Our approach is an adaptive per-SNR optimization, in which an optimal rotation for a given constellation and SNR value can be computed by a simple exhaustive search on the interval $[0,\pi/2]$.  Our rotations are applicable to non-uniform and uniform constellations alike, thereby improving the ability of non-uniformity to improve the cutoff rate.  \textcolor{black}{The cutoff rate has essentially served as a proxy for, and lower bounds, the CM capacity of the constellation.  The computational intractability of the CM capacity for the channel under consideration makes the cutoff rate a much more attractive objective function.}

Our rotations produce constellations which, while lacking full diversity, are ``locally'' fully diverse in the sense that every point is distinguishable from all of its closest neighbors by a single coordinate.  \textcolor{black}{We have thus defined a new notion of diversity, deemed \emph{local diversity}, which apparently suffices} to guarantee modest improvement in both cutoff rate and error performance over the number-theoretic methods of \cite{OV} \textcolor{black}{when the dimension is $n = 4$}.  By constructing a well-performing one-parameter subgroup of rotation matrices, we have added a degree of flexibility to the use of rotated constellations.  We have used this parameter to optimize with respect to SNR, but one may use it to optimize with respect to, for example, outer code rate as in the DVB-NGH standard \cite{DVBNGH}.  

Future work consists of generalizing our construction to create well-performing families of rotation matrices ${\bf Q}_n(t)$ for all $n$, not just those which are a power of $2$, \textcolor{black}{and additionally improving the performance of our scheme for dimension $n = 8$}.  The optimal parameter $t$ should also be investigated in connection with outer error-correcting codes.  \textcolor{black}{Secondly, we wish to study the decoding complexity of our rotation matrices, and perhaps construct some variants which offer provably simpler sphere decoding complexity.}  Lastly, we plan on investigating local diversity from an information-theoretic perspective to explain exactly why locally fully diverse constellations can outperform those which are globally fully diverse.  We hope that the resulting intuition can be applied to the study of other channels, for example MIMO channels and Rician channels, and not simply the SISO Rayleigh fading channel.  \textcolor{black}{We believe the robustness of our approach, especially the general independence of the numerical methods on the particular objective function we used, will prove that our ideas will be similarly useful in other settings.}

\bibliographystyle{ieee}
\bibliography{myrefs_new}

\begin{thebibliography}{10}

\bibitem{boutros}
J.~Boutros and E.~Viterbo,
\newblock ``Signal space diversity: A power- and bandwidth-efficient diversity
  technique for the {R}ayleigh fading channel'',
\newblock {\em IEEE Trans. on Inf. Theory}, vol. 44, no. 4, July 1998.

\bibitem{boutrousgood}
J.~Boutrous, E.~Viterbo, C.~Rastello, and J.-C. Belfiore,
\newblock ``Good lattice constellations for both {R}ayleigh fading and gaussian
  channels'',
\newblock {\em IEEE Trans. on Inf. Theory}, vol. 12, no. 2, March 1996.

\bibitem{OV}
F.~Oggier and E.~Viterbo,
\newblock ``Algebraic number theory and code design for {R}ayleigh fading
  channels'',
\newblock {\em Foundations and Trends in Communications and Information
  Theory}, vol. 1, no. 3, pp. 333--416, 2004.

\bibitem{massey}
J.~L. Massey,
\newblock ``Capacity, cutoff rate, and coding for a direct-detection optical
  channel'',
\newblock {\em IEEE Trans. on Comm.}, vol. 29, no. 11, pp. 1615--1621, November
  1981.

\bibitem{forney_ung}
G.~Forney and G.~Ungerboeck,
\newblock ``Modulation and coding for linear gaussian channels'',
\newblock {\em IEEE Trans. on Inf. Theory}, vol. 44, no. 6, pp. 2384--2415,
  October 1998.

\bibitem{fabregas}
A.~F\`abregas and G.~Caire,
\newblock ``Multidimensional coded modulation in block-fading channels'',
\newblock {\em IEEE Trans. on Inf. Theory}, vol. 54, no. 5, May 2008.

\bibitem{herath}
S.\ Herath, N.H.\ Tran, and T.\ Le-Ngoc,
\newblock ``Rotated {Multi}-{D} constellations in {R}ayleigh fading: Mutual
  information improvement and pragmatic approach for near-capacity performance
  in high-rate regions'',
\newblock {\em IEEE Trans. on Comm.}, vol. 60, no. 12, December 2012.

\bibitem{hero}
A.O. Hero and T.L. Marzetta,
\newblock ``Cutoff rate and signal design for the quasi-static
  {R}ayleigh-fading space-time channel'',
\newblock {\em IEEE Trans. on Inf. Theory}, vol. 47, no. 6, September 2001.

\bibitem{boulle2}
K.~Boull\'e and J.-C. Belfiore,
\newblock ``The cutoff rate of time correlated fading channels'',
\newblock {\em IEEE Trans. on Inf. Theory}, vol. 39, no. 2, March 1993.

\bibitem{edelman}
A.~Edelman, T.~Arias, and S.~Smith,
\newblock ``The geometry of algorithms with orthogonality constraints'',
\newblock {\em SIAM J. Matrix Anal. Appl.}, vol. 20, no. 2, pp. 303--353, 1998.

\bibitem{zoellner}
J.~Zoellner and N.~Loghin,
\newblock ``Optimization of high-order non-uniform {QAM} constellations'',
\newblock in {\em IEEE International Symposium on Broadband Multimedia Systems
  and Broadcasting (BMSB)}, 2013.

\bibitem{hossain}
J.~Hossain, A.~Alvarado, and L.~Szczecinski,
\newblock ``{BICM} transmission using non-uniform {QAM} constellations:
  Performance analysis and design'',
\newblock in {\em IEEE International Conference on Communications (ICC)}, 2010.

\bibitem{betts}
W.~Betts, A.~Calderbank, and R.~Laroia,
\newblock ``Performance of nonuniform constellations on the gaussian channel'',
\newblock {\em IEEE Trans. on Inf. Theory}, vol. 40, no. 5, September 1994.

\bibitem{zhongtakahara}
L.~Zhong, F.~Alajaji, and G.~Takahara,
\newblock ``Error analysis for nonuniform signaling over {R}ayleigh fading
  channels'',
\newblock {\em IEEE Trans. on Comm.}, vol. 53, no. 1, January 2005.

\bibitem{DVBNGH}
``Next generation broadcasting system to handheld, physical layer specification
  ({DVB}-{NGH})'',
\newblock
  \url{http://www.dvb.org/resources/public/standards/A160_DVB-NGH_Spec.pdf},
  November 2012.

\bibitem{lachlan}
M.~Lachlan and D.~Gomez-Barquero,
\newblock ``Modulation and coding for {ATSC} 3.0'',
\newblock in {\em IEEE Internation Symposium on Broadband Multimedia Systems
  and Broadcasting (BMSB)}, 2015.

\bibitem{DVBT2}
``Implementation guidelines for a second generation digital terrestrial
  television broadcasting system ({DVB}-{T}2)'', February 2012,
\newblock http://www.dvb.org/technology/standards.

\bibitem{big_broadcasting_book}
D.~G\'omez-Barquero,
\newblock {\em Next Generation Mobile Broadcasting},
\newblock Taylor \& Francis Group, LLC, 2013.

\bibitem{OBV}
F.~Oggier, J.-C.\ Belfiore, and E.~Viterbo,
\newblock ``Cyclic division algebras: A tool for space-time coding'',
\newblock {\em Foundations and Trends in Communications and Information
  Theory}, vol. 4, no. 1, pp. 1--95, 2007.

\bibitem{amin_rician}
A.~Sakzad, A.-L. Trautmann, and E.~Viterbo,
\newblock ``Cross-packing lattices for the {R}ician fading channel'',
\newblock in {\em IEEE Information Theory Workshop (ITW)}, 2015.

\bibitem{amin_full_diversity}
A.~Sakzad and E.~Viterbo,
\newblock ``Full diversity unitary precoded integer-forcing'',
\newblock {\em IEEE Trans. on Wireless Comm.}, vol. 14, no. 8, April 2015.

\bibitem{scam}
M.~Taherzadeh, H.~Nikopour, A.~Bayesteh, and H.~Baligh,
\newblock ``{SCMA} codebook design'',
\newblock in {\em IEEE Vehicular Technology Conference (VTC)}, 2014.

\bibitem{karpukrot}
D.~Karpuk and C.~Hollanti,
\newblock ``Rotating non-uniform and high-dimensional constellations using
  geodesic flow on {L}ie groups'',
\newblock in {\em IEEE International Conference on Communications (ICC)}, 2014.

\bibitem{karpukrot2}
D.~Karpuk and C.~Hollanti,
\newblock ``Multi-dimensional and non-uniform constellation optimization via
  the special orthogonal group'',
\newblock in {\em IEEE Information Theory Workshop (ITW)}, 2014.

\bibitem{caire}
A.~F\`abregas, A.~Martinez, and G.~Caire,
\newblock ``Bit-interleaved coded modulation'',
\newblock {\em Foundations and Trends in Communications and Information
  Theory}, vol. 5, no. 1-2, pp. 1--153, 2008.

\bibitem{Viterbo_rotations}
E.~Viterbo,
\newblock ``Optimal rotations for number field lattices'',
\newblock
  http://www1.tlc.polito.it/{\raise.17ex\hbox{$\scriptstyle\sim$}}viterbo/rotations/rotations.html.

\bibitem{hallie}
B.~Hall,
\newblock {\em {L}ie groups, Lie Algebras, and Representations: An Elementary
  Introduction},
\newblock Springer, 2003.

\end{thebibliography}

\end{document}